\newif \ifAuthorversion	\Authorversionfalse % for ACM only

\newif \ifDraft         \Draftfalse
\newif\ifFinal \Finalfalse
\newif \ifPreprint \Preprintfalse
\newif \ifReview \Reviewfalse

\Preprinttrue
\ifFinal\Draftfalse\fi

\newif \ifhyperlinks    \hyperlinkstrue
\ifPreprint
\documentclass[nonacm]{acmart} % no copyright notice
\usepackage{balance}
\else
\documentclass[\ifAuthorversion authorversion,\fi hyphens,sigconf,10pt\ifReview,nonacm,review\ifPreprint,authordraft\else,anonymous\fi\ifDraft,authordraft,timestamp,draft\fi\fi]{acmart}
\fi %!ifPreprint
\hypersetup{nolinks=false}

\usepackage{verbatim}           % for comment environment (eats up contents)
\newcommand{\code}[1]{\texttt{#1}}
\usepackage[final]{listings}
\usepackage[T1]{fontenc}
\usepackage[scaled=0.82]{beramono}

\usepackage{xspace}             % to fix spacing in macros producing words. Eg:
\setkeys{Gin}{keepaspectratio=true,clip=true,draft=false}
\graphicspath{{./imgs/}}

\ifDraft
  \usepackage{draftwatermark}
  \SetWatermarkLightness{0.85}
  \newcommand{\Comment}[1]{\textbf{\textsl{#1}}}
  \newenvironment{LongComment}[1] % multi-paragraph comment, argument is owner
    {\begingroup\par\noindent\slshape \textbf{\colorbox{yellow}{Begin Comment[#1]}}\par}
    {\par\noindent\textbf{\colorbox{yellow}{End Comment}}\endgroup\par}
  \newcommand{\FIXME}[1]{\textbf{\textsl{\colorbox{yellow}{FIXME:} #1}}}
  \newcommand{\TODO}[1]{\textbf{\textsl{\colorbox{yellow}{TODO:} #1}}}
\else
  \newcommand{\Comment}[1]{\relax}
  
  \newcommand{\FIXME}[1]{\relax}
  \newcommand{\TODO}[1]{\relax}
\fi

\usepackage{subcaption}

\ifhyperlinks\else
  \hypersetup{nolinks=true}
\fi

\newcommand{\utcs}{\(\mu\)TCs\xspace}
\newcommand{\defined}[1]{\mathsf{#1}}
\newcommand{\confu}{\defined{confidentiality{\text -}u}}
\newcommand{\getobj}{\defined{get{\text -}object}}

\begin{document}\sloppy

  %% Capitalisation of cross references
  \renewcommand{\sectionautorefname}{Section}
  \renewcommand{\subsectionautorefname}{Section}
  \renewcommand{\subsubsectionautorefname}{Section}
  \renewcommand{\appendixautorefname}{Appendix}
  \renewcommand{\Hfootnoteautorefname}{Footnote}
  %% Commands for index
  \newcommand{\Htextbf}[1]{\textbf{\hyperpage{#1}}}

\ifPreprint\else % remove ACM license for arXiv preprint
\ifReview\else
  \copyrightyear{2023}
  \setcopyright{acmlicensed}
  \acmConference[]{29th ACM Symposium on Operating Systems Principles,
    2023}{October 2023}{Koblenz, Germany}
  \acmDOI{...}
  \ifAuthorversion
    %% Hack to get same column break on author and ACM versions (adjust vspace as needed)
    \makeatletter
    \def\@formatdoi#1{\url{https://doi.org/#1}\vspace*{2ex}}
    \makeatother
  \fi
\fi
\fi %!ifPreprint

%%%%%%%%%%%
\ifDraft
Draft submission to SOSP'23. The CfP is here:
\url{https://sosp2023.mpi-sws.org/cfp.html}.

Limit is 12 pages plus references.

Ensure that there are no lines overflowing, especially in the listing
(indicated in draft mode by a black box at the margin).

Turn off Draft mode before submitting by setting  \verb,\Draftfalse, in the header.

Note, the Makefile producing the PDF suppresses all ``editor'',
``publisher'', ``issn'' and ``isbn'' strings from the .bib files, as
these just waste space and are generally not needed. In the rare
case where you want them (eg a journal paper) remove the respective
sed script from the Makefile.

Suppression of DOI strings is simpler to control: Simply add
\verb|\newcommand{\doi}[1]{\relax}| in the \LaTeX{} source and
they'll be suppressed. (In some document styles \verb|\doi| is
already defined, so you'll need to do \verb|\renewcommand|
instead. \verb|\providescommand| is unlikely to bring you joy.)

\clearpage
\addtocounter{page}{-1}
\makeatletter
\ACM@linecount\@ne\relax
\makeatother
\fi
%%%%%%%%%%%

\title{Proving the Absence of Microarchitectural Timing Channels}
%\subtitle{Subtitle Text, if any}
%\subtitlenote{Subtitle note}
\author{Scott Buckley}
\ifPreprint
\authornote{Joint lead authors. \copyright\ The owner/author(s) 2023}
\fi
\orcid{0000-0001-8810-9323}
\affiliation{\institution{UNSW Sydney}\country{Australia}}
\email{scott.j.h.buckley@gmail.com}

\author{Robert Sison}
\ifPreprint
\authornotemark[1]
\fi
\orcid{0000-0003-0313-9764}
\affiliation{\institution{UNSW Sydney}\country{Australia}}
\affiliation{\institution{University of Melbourne}\country{Australia}}
\email{r.sison@unsw.edu.au}

\author{Nils Wistoff}
\orcid{0000-0002-8683-8060}
\affiliation{\institution{ETH Zürich}\country{Switzerland}}
\email{nwistoff@iis.ee.ethz.ch}

\author{Curtis Millar}
\affiliation{\institution{UNSW Sydney}\country{Australia}}
\email{curtis.millar@gmail.com}

\author{Toby Murray}
\orcid{0000-0002-8271-0289}
\affiliation{\institution{University of Melbourne}\country{Australia}}
\email{toby.murray@unimelb.edu.au}

\author{Gerwin Klein}
\orcid{0000-0001-8883-0559}
\affiliation{\institution{Proofcraft}\country{Australia}}
\affiliation{\institution{UNSW Sydney}\country{Australia}}
\email{gerwin.klein@proofcraft.systems}

\author{Gernot Heiser}
\orcid{0000-0002-7069-0831} % this is Gernot's ORCID
\affiliation{\institution{UNSW Sydney}\country{Australia}}
\email{gernot@unsw.edu.au}

\begin{abstract}
  Microarchitectural timing channels are a major threat to computer
  security. A set of OS mechanisms called \emph{time protection} was
  recently proposed as a principled way of preventing information
  leakage through such channels and prototyped in the seL4
  microkernel. We formalise time protection and the underlying
  hardware mechanisms in a way that allows linking them to the
  information-flow proofs that showed the absence of storage channels
  in seL4.
\end{abstract}
\maketitle
\ifFinal
  \pagestyle{empty}
\fi

\section{Introduction}\label{s:intro}

\emph{Timing channels} bypass the operating system's (OS's) security
enforcement by leaking information through the timing of observable
events, such as the response time of a server or the observer's own
rate of progress. In the case of \emph{microarchitectural} timing
channels, this is achieved by manipulating hardware resources,
such as caches, TLB, branch predictors and prefetchers, that are
abstracted away by the \emph{instruction-set architecture} (ISA)~\citep{Ge_YCH_18}.

Like other forms of covert channels, timing channels break
\emph{confinement}, the ability to prevent a service from leaking a
client's secrets~\citep{Lampson_73}. Confinement is highly desirable
whenever sensitive data is processed by an untrusted service, such as
a web browser, or any third-party application installed on a computer
or mobile device. In principle, any code of sufficient complexity,
unless proved correct, must be suspected to contain bugs that can be
leveraged by an attacker into Trojans that leak secrets. Furthermore,
the Spectre attacks demonstrated the construction of a Trojan from
speculatively executed gadgets in innocent code, where the Trojan used
timing channels to leak secrets across security
boundaries~\citep{Kocher_HFGGHHLMPSY_19}.

In practice this means that it is infeasible to rule out the existence
of Trojans that leak information via covert channels, and we must
instead rely on the OS to enforce sufficient isolation. For \emph{storage
channels} this is a solved problem: the seL4
microkernel~\citep{Klein_EHACDEEKNSTW_09} was formally proved to be
free of storage channels~\citep{Murray_MBGK_12}. No such proof exists
for \emph{timing-channel freedom}.

However, there is hope: \citet{Ge_YCH_19} recently proposed \emph{time
protection}, a set of OS mechanisms designed fundamentally to \emph{prevent}
all possible leakage through microarchitectural timing channels
(hereafter referred to as \utcs) by making it impossible to pass information
through \emph{any} microarchitectural state.
These mechanisms included flushing on-core state on domain switch, partitioning
the OS's text and stack to occupy distinct per-domain colours in the off-core
memory caches, and padding time after flush operations up to their worst-case
latency.
\citeauthor{Ge_YCH_19} implemented time protection in an
experimental version of seL4 that supports only a separation kernel policy,
but also observed that -- even for this most restrictive of policies --
contemporary hardware lacked the mechanisms to enable the OS to completely
remove the channels, leaving a degree of vulnerability. More recently however,
\citet{Wistoff_SGBH_21,Wistoff_SGHB_23} added a suitable mechanism to flush all
on-core state with time padding up to a specified minimum execution latency, in
the form of a new instruction called \code{fence.t}, to
an open-source RISC-V processor core with minimal increase of
implementation complexity and no performance cost, making time
protection for on-core \utcs feasible.

This raises the intriguing possibility of formally
\emph{proving} the prevention of leakage through \utcs. Verification is
highly desirable for any code that is critical to security
enforcement. It is particularly important for the implementation of
time protection, as even logically correct code may lead to observable
timing variations.
To date, formal reasoning about such timing variations has been considered
extremely challenging and generally infeasible.

\citet{Heiser_KM_19} previously proposed that, as \utcs result from competing
access to hardware resources, and time protection strictly partitions
all such resources either spatially or temporally~\citep{Ge_YCH_19},
reasoning about timing channels can be transformed into
reasoning about storage channels.
However, we have found that making this link \emph{formally}, between reasoning
about storage access and their impact on time via the
microarchitecture, is highly non-trivial as it still
requires reasoning about the \emph{sequence} of accesses to that storage.

Moreover, information-flow properties are notoriously difficult to
reconcile with \emph{refinement}, i.e.\ proofs that a concrete
implementation satisfies an abstract specification, as refinement does
not generally preserve information flow. Traditionally, that challenge
is overcome
by removing any observable nondeterminism from the specification, as done
in the seL4 information-flow
proofs~\citep{Murray_MBGBSLGK_13}. However, this is not possible here, as
the exact behaviour of the microarchitecture is unspecified (for good reason).
We address this challenge by an over-approximation of accessed storage
that supports reasoning about timing channels that is preserved under
refinement.

This over-approximation is a significant departure from the existing
verification approach of seL4, creating a new challenge of integrating
with the a huge, organically grown proof base not designed with such
changes in mind (see \autoref{s:proof-reuse}).
In tackling this challenge, we base our work on the original time protection
OS mechanisms of~\citep{Ge_YCH_19}, as reimplemented by us on RISC-V to take
advantage of the new \code{fence.t} instruction of~\citet{Wistoff_SGBH_21}.
This presents a starting point that supports only separation-kernel policies
that enforce complete isolation between domains;
we found the challenges in integrating time protection into the seL4 proof base
even for the separation kernel case to be significant enough that we must
consider non--separation-kernel policies as future work.

We make the following contributions:
\begin{itemize}
\item a new abstraction of \emph{touched addresses} as an over-approximation of address traces that enables information-flow reasoning (\autoref{s:unknown-trace}) in a manner we believe is amenable to refinement (\autoref{s:refinement-directions});
\item a reasoning framework for time protection that utilises the above abstraction, and is \emph{external} to the existing seL4 proof system (\autoref{s:extension});
crucially, this includes a proof that \textbf{if} we have a secure system model
that tracks the touched addresses,
we can produce a proof of time protection for that system, as long as well-defined system and hardware requirements are met;
\item a formalisation of the aforementioned hardware requirements --
the hardware-software contract --
that supports verification of time protection (\autoref{s:hardware});
\item changes to the kernel implementation
as a result of our formal reasoning, and a report on our current progress
in the repair of the seL4 proofs to return them to a fully-proved state
including an investigation of the future work needed to integrate the external
time-protection model with the seL4 proofs
(\autoref{s:integration});
\item experimental validation that time-protection implementation remains effective despite these changes (\autoref{s:benchmarks}).
\end{itemize}

Here, as in \citet{Ge_YCH_19}, we limit our scope to assume a single-core
system with no hardware-level multithreading -- a necessary limitation for us
given that the multicore verification of seL4 is an unfinished area of active
research.

\section{Background}\label{s:background}

\subsection{Covert channels}

A \emph{covert channel} is an information flow that uses a mechanism
not intended for information transfer~\citep{Lampson_73}. As such it
can be used to violate a system's security policy, by leaking
information across security boundaries that do not authorise such
information flow. Some covert channels may serve as \emph{side
  channels}, if an attacker is able to extract a secret from a
security domain as a side effect of an innocent execution of the
domain that holds the secret. In contrast, a covert channel that is
not a side channel requires insider help, e.g.\ a \emph{Trojan} that
colludes with the attacker (also referred as a \emph{spy}) to actively
leak information. Such a covert channel represents the worst case of
leakage, and is what we concern ourselves in this work.

Covert channels are further classified by the mechanisms they
exploit. Physical channels, such as temperature~\citep{Murdoch_06,Masti_RRMTC_15}, power
draw~\citep{Kocher_JJ_99} or electromagnetic~\citep{Quisquater_Samyde_01,Genkin_PPT_15} or
acoustic~\citep{Backes_DGPS_10,Genkin_ST_14} emanation, generally require physical access or at
least proximity, these are out of scope of this work. In contrast, a
channel that can be used by a (remotely controlled) unprivileged spy
program for unauthorised access to protected secrets is a primary OS
concern. Traditionally these are classified as either \emph{storage}
or \emph{timing channels}, depending on whether or not their
exploitation requires the measurement of timing of
events~\citep{Schaefer_GLS_77,DoD_85:orange}, although \citet{Wray_91}
notes that this distinction is more of an approach to exploitation
than fundamental.

Nevertheless, storage channels are considered an easier problem than
timing channels, due to the difficulty of reasoning about (or
preventing) minor variation in execution timing. Specifically, the
information-flow proofs of seL4 \citep{Murray_MBGBSLGK_13} prove the
absence of leakage through storage channels but are quiet on timing
channels.

\subsection{Microarchitectural timing channels}

Timing channels can have various causes, such as data-dependent
control flow in crypto algorithms~\citep{Molnar_PSW_06}. The OS cannot
be held responsible for such algorithmic channels. In contrast,
microarchitectural channels exploit hardware resources, and
only the OS has full control over hardware and the OS is therefore responsible
to closing such channels.

\utcs exploit hardware that is functionally transparent and as such
mostly or completely abstracted away by the hardware-software
contract, the instruction-set architecture (ISA). The purpose of such
hardware is improving average-case performance, which implies that its
presence (and state) can be sensed by observing execution
speed. Examples are the CPU I- and D-caches, TLBs, branch predictors
and prefetchers~\citep{Ge_YCH_18}. These hardware resources hold state
that depends on recent execution history and is used to speed up
future operations, based on temporal or spatial locality.

Exploiting these channels involves probing the footprint left by past
execution on the hardware state. A typical approach is
\emph{prime-and-probe}~\citep{Osvik_ST_06, Percival_05}: The spy
\emph{primes} the hardware, e.g.\ by reading a large enough buffer to ensure
that the L1 D-cache only holds spy data. It then waits for the Trojan
to execute, leaving its footprint in the cache. The spy then \emph{probes} the
cache, again by accessing the large buffer to touch every cache line,
while observing access latencies. Fast access means that the spy's
original data is still cached, while slow access means that the
corresponding cache line has been replaced by the Trojan, thus
conveying information from Trojan to spy. Similarly, the I-cache can
be probed by using jumps instead of loads, and the same basic approach
can be used to probe the TLB, branch-predictor and prefetcher state.

\subsection{Time protection}

\citet{Ge_YCH_19} proposed time protection (as the temporal equivalent
of memory protection) as a principled way of preventing timing
channels. They postulate that timing channels are eliminated if all
potentially shared hardware resources are either spatially or
temporally partitioned.

Spatial partitioning applies to off-core caches, which are physically
addressed, and hence can be partitioned by the OS using page
colouring~\citep{Kessler_Hill_92, Lynch_BF_92, Liedtke_IJ_97}: the OS
prevents competing accesses to the same cache lines by allocating
physical memory so that the cache footprints of different domains
cannot overlap.

On-core state is accessed by virtual address, which is out of control
of the OS, and therefore must be temporarily
partitioned, i.e.\ securely multiplexed~\citep{Ge_YCH_19}. This means
that all such hardware resources must be reset to a defined state when
handing the hardware to a different security domain.

As temporal partitioning cannot help if hardware is accessed
by different security domains concurrently, this means hyperthreading
must be disabled or all of a core's hyperthreads must belong to the
same domain. Furthermore, as the reset latency may depend on the hardware
state (at least in the case of data caches that require writing back
dirty data), it must be padded to the worst-case latency to prevent
the reset time from becoming a channel.

Colouring \emph{all} memory is challenging, given that the kernel needs to
execute instructions and access data. \citet{Ge_YCH_19} address this
by \emph{cloning} the whole kernel, so each domain has its own kernel
image that handles system calls, each with its own text and data
sections.

However, there is a small amount of shared global kernel data that cannot be
cloned as it is required to coordinate kernels on a domain switch, via whose
footprint in the off-core caches \citet{Ge_YCH_19} identified the
opportunity for an inter-domain timing channel.
\citeauthor{Ge_YCH_19} attempted
to make accesses to this data deterministic by pre-fetching it into
the caches on a domain switch; in this paper, we instead model the treatment of
these accesses via a hardware primitive that allows a targeted flush of this
state from the off-core caches, as claimed to be needed by
\citet{Sison_BMKH_23} -- we discuss our findings and recommendations for
future work on this issue in \autoref{s:prefetch}.

\citet{Ge_YCH_19} do not deal with channels resulting from
bandwidth-limited shared interconnects~\citep{Hu_91, Wu_XW_12}, and
instead assume that all cores are co-scheduled to the same security
domain. They also leave channels resulting from off-chip state, such
as memory controllers~\citep{Wang_FS_14, Deutsch_YBDEY_22}, out of
scope. We limit our scope in the same way, and for now explicitly
assume single-core operation.

Within these constraints \citet{Ge_YCH_19} demonstrated that time
protection can be implemented at low overhead and is largely effective
in eliminating timing channels. However, they also found that present
main-stream x86 and Arm processors contain microarchitectural state
that cannot be reset by any architected mechanism, resulting in
residual state (in branch predictors and prefetchers) that the OS is
unable to close.

\citet{Wistoff_SGBH_21, Wistoff_SGHB_23} have meanwhile demonstrated a
new instruction, \code{fence.t}, that performs a full, deterministic reset of all
non-architected state (which they call \emph{microreset}) and enforces
a specified minimum execution latency (``time padding''). They
implemented \code{fence.t} in the Arianne (aka.\ CVA-6),  an open-source
RISC-V core. They find that
\code{fence.t} eliminates all known \utcs on their processor. They
also show that \code{fence.t} is cheap in terms of silicon overhead
(0.4\%) as well as execution time (no latency added over that of the
L1-D-cache flush).

\subsection{seL4}

seL4 is a microkernel with a comprehensive formal verification story,
covering functional correctness of the implementation down to binary
code, and proofs that the kernel is able to enforce integrity,
availability and confidentiality~\citep{Klein_AEMSKH_14}. Originally
done for 32-bit Arm processors, the proofs have meanwhile been
extended to the 64-bit RISC-V architecture
(RV64)~\citep{rv64_infoflow}.  The information-flow proofs
establishing confidentiality exclude storage channels through architected state
the kernel is aware of, but explicitly do not address timing channels.

\citet{Ge_YCH_19} implemented time protection in an experimental
version of seL4, which added kernel cloning. Each thread in the system
is associated with a kernel image, and when, during normal execution
of the kernel, the scheduler selects a thread bound to a different
kernel, the kernel performs the time-protection operations before
switching the kernel page-table pointer (which switches the kernel
image).

In contrast, seL4's information-flow proofs are based on a
\emph{domain scheduler} which implements strict time partitioning
(fixed, round-robin time slices for each domain), with a secondary
scheduler in each domain responsible for scheduling inside a time
partition. Our work aims to leverage the existing information-flow
proofs and as such is based on this domain scheduler.

\subsection{Refinement}

seL4's verification uses refinement~\citep{deRoever_Engelhardt:DR}: A
concrete model (implementation) is proved to only exhibit a subset of
behaviours of an abstract model (specification). In seL4's case there
is a hierarchy of models:
\begin{enumerate}
\item Abstract statements of security enforcement (integrity and
  confidentiality);
\item an access-control model that lumps fine-grained seL4
  capabilities into classes of access;
\item the (operational) abstract specification of the kernel, describing
  individual system calls, their arguments and how they act on the
  state of the abstract system model;
\item an executable specification (originally derived from a Haskell
  implementation);
\item the implementation (C code formalised in the theorem prover
  through a parser written in Isabelle/HOL)~\citep{Tuch_KN_07};
\item the executable binary code (formalised through an ISA
  specification in the HOL4 prover~\citep{Fox_Myreen_10}).
\end{enumerate}

Most of these proofs are performed by interactive theorem proving
(ITP) in the Isabelle/HOL proof assistant~\citep{Nipkow_PW:Isabelle}
(the exception is the last refinement step, which uses an automated tool chain).
Confidentiality is proved by formal information-flow
reasoning~\citep{Murray_MBGBSLGK_13}. Intuitively, there is no flow of
information from a secret-holding ``High'' domain to another ``Low''
domain, if Low is not influenced by the state of High. Specifically,
assume we compare two execution traces of Low which start from the
same global system state, except for the value of a secret held by
High. If the two traces are indistinguishable then there is no
information flow from High to Low.

An inherent challenge here is that refinement generally does not preserve
information flow confinement. To illustrate this, consider a specification
that states that a function intended to be a random-number generator
returns any integer; the function can be called by anyone without requiring
specific privilege. The exact functionality is left unspecified via
non-determinism. Any implementation that returns an integer is a refinement
of that specification, even if it returns the value of a secret
variable, which obviously violates confidentiality.

One way to preserve reasoning about confidentiality through refinement is to
make all observable behaviour deterministic in the specification -- this is the
approach chosen in seL4's confidentiality proofs. However, this is only possible
if all information on which decisions are made can be made visible in the
specification. This is not always feasible. In particular for our purposes, we
will be interested in the sequence of accesses to memory, which in turn
determine the impact on the microarchitectural state that causes timing
channels. This exact sequence is not necessarily available.

\section{Time-Protection Verification Challenges}\label{s:challenges}

\subsection{The unknown-trace problem}\label{s:unknown-trace}

As mentioned above, information-flow proofs of confidentiality enforcement
reason about execution traces of programs. However, the very nature of
the microarchitecture makes it infeasible to have exact traces, and
therefore to compare arbitrary traces precisely. This would require a
precise model of the microarchitecture, but details
of the microarchitecture are intentionally left unspecified by the
hardware-software contract, to enable portability of code and retain
hardware manufacturers' freedom to optimise. For example, out-of-order
processors will reorder instructions, which changes memory-access
traces. Similarly, cache-line replacement policies affect cache
residency, which also affects memory traces.

Furthermore, even if there was a sufficiently complete model of the
hardware to make traces deterministic, this would be far too complex
for formal reasoning.

So we have to accept the \emph{unknown-trace problem}, and thus our
inability to reason about precise traces.

We address this by an over-approximation: We replace an exact memory
trace by the \emph{touched address} (TA) set, the set of all memory
locations that we predict to be \emph{potentially} accessed during the
execution of a security domain.
Some examples of typical over-approximations in these predictions are (1)
including the addresses occupied by the entirety of a given kernel object when
only part of it is accessed or (2) including every object in a table when only
one of them is accessed.
Key to our approach will be that the extent of these
overapproximations is irrelevant as long as the proofs maintain the invariant
that the TA set remains a subset of the appropriate partition
(see \autoref{s:tp-prop-ext}, \autoref{thm:ta-inv}).
It will then fall within the scope of certain ``no-fail'' obligations
in the refinement proofs (see \autoref{s:ta-coverage}) to enforce our
abstract specification-level assertions that the actual addresses accessed
\emph{always} remain a subset of the predicted TA set.

This approach enables us to prove a stronger property than the
standard equality of traces: We quantify over \emph{all} traces of interactions with addresses
that lie within this over-approximated set -- in our model, this will be all
sequences of reads, writes, and off-core flushes that impact the
microarchitectural state relevant to addresses within that set (see
\autoref{s:the-model}).
This in turn supports proving a
formal notion of OS-enforced time protection that is meaningful at any level of
abstraction of the OS. Furthermore, this formalisation of time protection is
compatible with refinement, as it allows reasoning about the impact of
each stage of refinement on the TA set.

\subsection{Proof reuse}\label{s:proof-reuse}

Interactive theorem proving, as used for seL4, is a very labour-intensive
process.  The total proof base of seL4 by now comprises well over a
million lines of proof script~\citep{github:l4v}, likely the
largest interactive bodies of proof ever constructed. Hence, when
attempting to verify any variant of seL4, such as one that supports
time protection, the ability to re-use most of the existing proof base
is of critical importance.

Predating the concept by over a decade, naturally none of the seL4
specifications or proofs were written with even a \emph{thought} of one day
attempting to use them to formalise and prove time protection. Indeed, the seL4
authors initially considered proving absence of timing channels
infeasible~\citep{Klein_MGSW_11}.

The upshot is that the changes needed to verify time protection in
seL4 must be minimally invasive on the existing proof base. This needs careful
proof engineering and is
especially true for incorporating the concept of touched addresses, as
it alters the essence of the seL4 specifications and proofs: it potentially
introduces state change into many parts of the specification that previously
were read-only.

In the next section we present an extension model for reasoning about
time protection that is separate to the existing seL4 system proofs. We will discuss integration of the models in \autoref{s:integration}.

\section{The Time Protection Extension}\label{s:extension}
In this section we will discuss the design of an external system model
for proving time-protection properties about seL4's abstract specification.
First we explain why this external model can guarantee the absence of
\utcs in the host model (i.e.\ the seL4 model) despite being entirely external to it.
We then describe the extension itself
and the security property we have proved will hold, as long as
certain system and security requirements hold about the original model.
We defer a discussion of the hardware assumptions until \autoref{s:hardware}.

\subsection{Why an extension is sufficient to capture \utcs}

We follow the same justification as \citet{Sison_BMKH_23} for defining time
protection as equivalences on the microarchitectural state abstractions
rather than modelling the activity of the spy directly; this allows us
to justify modelling the microarchitecture in an extension of the model rather
than integrating it directly into the original model.

A high level view of the models in question might look like the following.
If we consider the existing system model from seL4's proof base, and describe this
as a transition system $T_a$, this system is aware of what we call \emph{abstract state} --
it is not aware of low-level details about hardware, and certainly has no knowledge
of microarchitectural state.
Our only modification to $T_a$ is to have it track a touched-address set
(as described in \autoref{s:unknown-trace};
details relegated to \autoref{s:integration}).

Although our goal is to reason in detail about microarchitectural state and its interaction with other hardware and time,
it is not practical to add that state directly to
seL4's proof base ($T_a$).
This would require us to model all of that state's impact on the original state, impacting all existing proofs.
As outlined in \autoref{s:proof-reuse} we  preserve existing proofs
by creating an external system model $T_\mu$ in which to model microarchitectural state transitions.

In creating this new transition system $T_\mu$, we seek to represent real-world flows of information
between microarchitectural state and the higher-level hardware state that is represented in seL4's abstract state
(hereafter referred to as $s_\mu$ and $s_a$ respectively).
In this conceptualisation, behaviour from a Trojan leaving a fingerprint on $s_\mu$ is
then considered a flow from $s_a$ to $s_\mu$,
and a spy observing this fingerprint in $s_\mu$ is considered a flow from $s_\mu$ to $s_a$.

With the transitions in $T_\mu$ modeled from the transitions in $T_a$,
with microarchitectural consequences in $s_\mu$
following from transformations to $s_a$,
we can faithfully model any information flows caused by the Trojan
from $s_a$ to $s_\mu$.

In modelling $T_\mu$ as an extension of $T_a$, we have chosen
not to model the flows induced by the spy back from $s_\mu$ to $s_a$; indeed,
with seL4's model $T_a$ never being aware of microarchitectural state, the combined model is not directly able to represent
a spy observing microarchitectural state and making decisions based on what is observed.

Instead we argue that, for proving the absence of \utcs,
it is sufficient to prove that any microarchitectural state
accessible to a spy contains no secrets while the spy is executing;
this is because any spy-induced flow of information from $s_\mu$ back to $s_a$
must follow from information already in $s_\mu$, which is thus
capturable by an information-flow property on $T_\mu$.

\subsection{The model of microarchitectural state}\label{s:the-model}
We achieve the above aim by adapting the methods presented by \citet{Sison_BMKH_23}: include
\emph{flushable state}, \emph{partitionable state}, and \emph{time} in $s_\mu$.
From here on we use the term \emph{flushable} to refer to \emph{temporally partitionable} state, and
refer to spatially partitionable state as simply \emph{partitionable}.

We enumerate different low-level hardware interaction units, and model how these interactions
affect microarchitectural state:
\begin{description}
      \item[Reads:] given some (virtual and physical) address, a memory read operation will have some
        underdefined impact on the flushable state, and will impact any relevant areas of the partitionable state.
        For example with coloured off-core caches, this will have an impact on the current colour of the L2 cache.
  \item [Writes:] these will have similarly undefined impacts on both the flushable and partitionable states.
        The in-model behaviours of read and write impacts on the microarchitectural are implemented identically,
        but are separated so that somebody using this model could represent the different microarchitectural
        effects of read and write operations.
  \item [On-core flush:] represents the \code{fence.t} primitive presented in \citet{Wistoff_SGBH_21},
        which in our model returns all flushable state to a predefined value.
  \item [Targeted off-core flush:] allows some section of the partitionable state to be flushed.
        The semantics here return some parts of the partitionable state to a predefined value, specifically
        the state relating to some set of addresses provided, as well as any addresses deemed to \emph{collide}
        in the partitionable state (i.e.\ addresses that map to the
        same cache set). This would flush the entire cache sets that
        may hold data addressed by a particular physical addresses.
  \item [Pad to time:] simply advances the time to a given constant, assuming it is in the future.
        This represents a hardware padding operation, and does not have any impact on flushable or partitionable
        microarchitectural state, apart from the clock itself.
\end{description}

The operations for reading, writing, and targeted off-core flushing are all parameterised by addresses (both virtual
and physical). We consider an operation to \emph{adhere to} a touched addresses set if the address(es) used by the operation
are contained within the touched addresses set.
We consider a full hardware interaction trace to be a sequence of these operations, and we consider a trace
to adhere to a touched addresses set if all of its component operations adhere.

To model the microarchitectural transformations associated with an abstract state transition $t_a$ in $T_a$,
we choose a trace that adheres to the touched addresses set associated with $t_a$.
We assume an underdefined \emph{trace selector} function that, given a touched addresses set, produces some hardware
interaction trace that adheres to the set.
For example, if the TA set is $\{1,2\}$ the selector may return
$[\mathtt{Read}\ 1, \mathtt{Write}\ 2, \mathtt{OffCoreFlush}\ \{1,2\}, \mathtt{OnCoreFlush}, \mathtt{Pad}\ t]$,
or its reorderings or extensions swapping any of the addresses with anything in
$\{1,2\}$ and any time $t$.

Given that our proofs have no knowledge of which trace is selected, except that we know that the selected trace
adheres to the touched addresses set, our proofs must allow that any possible trace could be selected, effectively
quantifying over all possible traces.

Although we quantify over all traces whose accesses stay in the TA set,
we place a crucial \emph{dependency}
restriction on the underdefined trace selector function: the
trace it selects must depend only on microarchitectural state \emph{visible} to the appropriate observer during
a particular transition.
Consider the following example, in which a system contains two security domains $A$ and $B$.
At any given time, either $A$ or $B$ is the currently-executing domain -- when the kernel is
executing, it is either doing so on behalf of one of these domains, or it is performing a domain-switch
between these domains.

During some transition $t_b$, if $B$ is the currently-executing domain, then from the perspective of $A$,
the transition could branch on any number of secrets belonging to $B$, so we have no idea what their trace will look like.
We assume that $t_b$ will encode secrets in all of the flushable state, as well as any of the partitionable state allowed
by the transition's touched addresses set. We know nothing about the time at the start or end of this transition.
To know that no secrets are learned by $A$, we need to know that $A$ cannot observe any of the microarchitectural state
affected by this transition.
We consider $A$ (as a non-executing domain) to be able to observe only the parts of partitionable state that are assigned
to $A$, excluding parts of this state that can be affected by kernel global data (which $B$'s transitions could impact).
We therefore only need to know that $B$'s touched addresses set contains no addresses in $A$'s partition of the partitionable
state, except for those in the kernel global data.

From the perspective of $B$, however, the flushable state is considered observable, as well as $B$'s partition of the
partitionable state, \emph{and} any part of the partitionable state affected by kernel global data.
So, from $B$'s perspective, the transition $t_b$ will affect plenty of visible microarchitectural state.
To preserve confidentiality in this instance, we need to know that the details of the transition are not dependent
on any of $A$'s secrets.
In particular, aside from the kernel global data,
we know the transition will only affect parts of $B$'s partitionable state,
as the traces will be induced from a TA set relevant to addresses wholly
within $B$'s partition; however, we still need to know that the
\emph{choice} of how that state is affected -- for example,
the order of $B$'s sequence of accesses to its own partition --
is not affected by any secrets from $A$.
To enforce this, we only need to know that the trace selector function did not consider any of $A$'s secrets when choosing
a particular hardware interaction trace to return.

This boils down to the following assertion: the trace selector function can only consider microarchitectural state
visible to the \emph{currently-executing} domain when making its decision on which adherent trace to select.
In real-world terms, this means that we are assuming that during some execution, the hardware trace is only affected
by microarchitectural state that is modeled as visible to the currently-executing domain.
This means we must carefully choose our definition of \emph{observable} to accurately reflect reality.

We define the parts of microarchitectural state observable to the currently-executing domain to be:
\begin{itemize}
  \item the entirety of the flushable state;
  \item the portion of the partitionable state assigned to the observing domain, plus
  \item any parts of the partitionable state colliding with kernel global data;
  \item the exact wall-clock time.
\end{itemize}

Therefore, the only parts of the state that the trace selector \emph{cannot} consider when selecting a trace
are the parts of the partitionable state that are assigned to other domains \emph{and} not colliding with
kernel global data addresses.
Effectively, this formalisation leans on \emph{hardware partitioning} being correctly enforced by hardware -- in terms of
memory caches, this means that off-core caches must be correctly coloured.
As long as this partitioning is correctly implemented, we can prove the absence of \utcs, when
appropriate time protection measures are implemented.

\subsection{Time protection measures in the model}
As in previous work \citep{Sison_BMKH_23}, the time protection mechanisms we model are as follows.
\begin{itemize}
  \item Correctly partitioning the partitionable state.
  \item Flushing the flushable state during a domain-switch. This models a
    flush of the entirety of \emph{all on-core microarchitectural state}, which
    we consider infeasible to partition -- including the L1-D/I caches, TLBs
    and branch predictors. Note in particular this is not just restricted to
    state relevant to addresses in the TA set.
  \item Selectively flushing parts of the partitionable cache during domain-switch.
    This is used to model the flushing of
    (statically determined) addresses of kernel global data from the
    \emph{off-core} i.e.~L2 cache (and L3, if it exists).
    Note this is only relevant to the TA set to the
    extent that we automatically include those globals in the TA set.
  \item Padding the domain-switch sequence to an appropriate
    worst-case execution time (WCET).
\end{itemize}

The model described in this section allows for the flow of secrets into the flushable state (for example on-core caches),
but these secrets are removed by a flush before execution is passed to any other domain.
The model allows for the flow of secrets into some portions of the partitionable state belonging to other domains (for example
off-core caches, via kernel global data), but this section of the partitionable state is also flushed before execution
is passed to any other domain.
The model allows for the flow of secrets into the rest of the currently-executing domain's portion of the partitionable
state, but this state is never observable by another domain.
Finally, the model allows for the flow of secrets into the clock itself, as the amount of time taken for a domain-switch
sequence to complete can be dependent upon domain secrets, but these secrets are removed when the clock is padded to
a domain-switch WCET.

To facilitate proofs for all of the above, we require only that the following property of the touched addresses
set holds for every transition:
\emph{The touched addresses set for each transition contains no addresses outside of the currently-executing domain's portion of the
partitionable state, except for addresses affected by kernel global data.}
For all kernel models that meet these requirements, the above claims are formally verified in Isabelle/HOL.

\subsection{Enforcing the presence of the time protection mechanisms}\label{s:enforcing}
Thus far we have discussed how to examine effects on the microarchitectural state that become
the medium for \utcs.
There is also the question of how to prove that time protection mechanisms are properly implemented.

When working with an abstract specification that has no knowledge of microarchitectural state,
there is no way to examine the semantics of the model to ensure that it is performing
time protection strategies in the right way or at the right times.

Instead, we ask the user of our model to specify which specific transitions implement
time protection mechanisms, and our microarchitectural model then simulates these mechanisms by executing
the primitive flushing and padding operations that make up the time protection implementation.

Specifically we request that a transition that is identified as a domain-switch transition to be
separable into four subtransitions: an \emph{old clean} transition, an optional \emph{dirty} transition,
a \emph{mechanism} transition, and a \emph{new clean} transition.
\begin{itemize}
  \item The \emph{old clean} sub-transition needs to meet the same transition requirements as any
        other transition in the system, via its touched addresses set.
  \item The \emph{dirty} sub-transition (not used in our integration with seL4) might access addresses
        belonging to both the \emph{previous} and \emph{next} domains (and therefore affect partitionable
        state belonging to both), but must have a hardware trace that is determinised only by information
        known to both the previous and next domains.
  \item The \emph{mechanism} sub-transition must \emph{only} perform a specific set of primitive microarchitectural
        operations: flushing of the on-core caches, any targeted flushing of off-core caches, and padding to a set time.
  \item The \emph{new clean} sub-transition is similar to the \emph{old clean} transition, but is now considered
        to be operating under the new domain's execution, so must only access addresses belonging to the new
        domain and in a sequence decided only by secrets visible to the new domain.
\end{itemize}

Our model executes all transitions as described above, and we then prove
that a confidentiality property capturing time protection
holds of this model as follows.

\subsection{Time protection property of extension}\label{s:tp-prop-ext}

First, we define the
TA set to be the union of all virtual addresses associated with all
\emph{user-controlled kernel objects} --
those allocated by a user program by retyping untyped memory --
that the kernel has retrieved so far since the last domain switch, as tracked
by calls to the seL4 abstract specification's internal $\getobj$ function.

Our new time-protection--relevant proofs, of
(1) information-flow security and (2) refinement, need to refer to the
TA set. We therefore add it as a \emph{ghost state} field
in the seL4 abstract specification -- that is, state that is intended to assist
proofs but that should not impact the functional semantics of the actual system.
(We discuss the mechanics of ensuring the TA set consequently does not impact
proof reuse further in \autoref{s:minimise-proof-impact}.)

In the seL4 information-flow security proofs,
we can compare
the values in the TA set with policy information determining which kernel
resources belong to which domains.
For seL4, we lean on an existing invariant proved about seL4 that
the page table maintains the mappings to all user-controlled
kernel objects at all times, regardless of which domain they belong to,
and that these will be unaffected by a domain switch.
Thus, to ensure kernel accesses impact only the
desired parts of the partitionable microarchitectural state,
which is indexed by physical -- not virtual -- address,
we must assert the kernel
maintains the following invariant:

\begin{proposition} [Partition subset invariant]
\label{thm:ta-inv}
The physical translations of all addresses
in the TA set, according to the page table,
form a subset of
the union of physical addresses that
reside in the currently running user domain's partition.
\end{proposition}

The main theorem we have proved of the system extension is as follows:
\begin{theorem} [Time-protection extension preserves confidentiality]
If $\confu$ holds for some transition system
that consists only of sequences of
(1) operations that obey the partition subset invariant and
(2) a correctly implemented domain switch sequence,
then $\confu_\mu$ holds of its time-protection extension.
\end{theorem}
Here, $\confu$ applied to the original system
is a version of seL4's original confidentiality property
\citep{Murray_MBGBSLGK_13}
modified to enforce an extra equivalence between the values of the TA set;
note this constitutes an extra security proof obligation for seL4.
Then, $\confu_\mu$ as applied to the extended system is the same but extended
to assert extra equivalences to \emph{enforce no secrets flow into
microarchitectural state fields observable to the currently executing domain}
as specified at the end of \autoref{s:the-model}.

\section{Hardware Requirements}\label{s:hardware}

The key hardware features necessary, as laid out by \citet{Sison_BMKH_23}, are
that primitives exist for flushing the on-core caches, that the
off-core caches are partitionable through cache colouring, and finally that a
primitive exists for waiting until a deterministic amount of time until after
the arrival of the timer interrupt that triggers domain-switch.
These now exist in the form of the \code{fence.t} instruction on the CVA-6 RISC-V core~\citep{Wistoff_SGBH_21,Wistoff_SGHB_23}.

\citet{Ge_YCH_19} used pre-fetching of shared kernel data to give it a
deterministic cache state, which we find to be insufficient and we
would need to flush any shared kernel data from the cache completely
(see \autoref{s:prefetch}). In order to avoid the prohibitive cost of
a full cache flush, this would require a per-address flush primitive.

However, with time protection tied to seL4's domain scheduler,
practically all of \citeauthor{Ge_YCH_19}'s shared kernel data can be
partitioned, and the tiny remainder (a link to the next domain to be
scheduled) is accessed deterministically, so this complication can be
completely avoided (for now, at least).

We obviously require that any flush operations used have a known WCET.
Moreover, we require that the hardware's
timing behaviour is reliable enough to establish the WCET of any
interrupt or system call triggered just before the timer interrupt
that triggers the domain switch. Note that \citet{Ge_YCH_19} defer operations
that occur within a short window before the timer tick until the
domain is scheduled next.

We improve on the model of \citet{Sison_BMKH_23} by allowing the "cachedness"
of a given address in the microarchitecture to be expressed as more than just a
boolean ("is resident" vs "is not resident"), to allow for data to be
cached at different levels in the cache hierarchy.

Moreover, we have discovered a number of requirements not previously described
in \citet{Sison_BMKH_23}, that more precisely express
on which parts of the microarchitecture the running time of these
operations depends.
For example, we formalise that the time taken by an on-core microarchitectural
flush depends only on other on-core microarchitectural state; moreover, we
require that the time taken by the flush of an address from an off-core cache
depends only of the cachedness state in that cache of other addresses of that
\emph{colour} -- the set of addresses whose bits in the overlap of the cache
index bits and page bits share a common value.
In a similar vein, we require that the access time to a particular address as
determined by the off-core cache should depend only on the cachedness state
of addresses in its \emph{collision set}. i.e.\ the set of addresses
that compete for the same cache space.

\section{Integration with seL4}\label{s:integration}
The existing proof base for seL4's security and functional correctness proofs is
a significant piece of engineering.
Here we explain how
we implement the touched-addresses abstraction
so as to minimise the impacts on seL4's existing proofs;  we then
discuss this impact.
We then describe changes we made to the design of the kernel's
implementation of time protection both enabled by, and to facilitate, our
formal reasoning.
Finally, we explain how we expect our abstractions
to enable seL4's refinement to be extended
to make the preservation of a time protection property possible.

\subsection{Minimising the proof impact}\label{s:minimise-proof-impact}

A significant portion of seL4's functional correctness proofs relies on
identifying sections of kernel execution that have no impact on the
model's state; this is critical to allow
the application of proof rules to be reordered or manipulated in various ways.
This could include lengthy sequences of complex executions, as long as memory is only
read, and never written to (or written to, but without changing values).

Unfortunately,
the tracking of touched addresses still constitutes an update
to the model state on every mere memory access.
To overcome this problem, we adapt the former ``no effect on state''
characterisation of executions
to express ``no effect on state other than the TA set'', allowing
reuse of the many existing proofs that have nothing to do with the TA set.

\subsection{Ensuring TA tracking is comprehensive}\label{s:ta-coverage}

To guarantee that accesses are adequately tracked by the TA set, we
leverage the failure-tracking mechanisms of the state transformation monads
used to model seL4's formal semantics:
We model memory accesses (including reads) to \emph{fail} if attempting to access any
address that is not already present in the state's touched-addresses set.
As long as the TA set has no way of losing addresses during a transition,
we can examine the touched addresses set at the end of that transition to gain an upper bound
on the set of addresses accessed during the transaction.

There are a few things to note here. Firstly, how do we know that executions do not fail?
This is achieved in the existing proof base during refinement, and is not a section of the proofs
that we have done significant work in updating. We do know that once ``no-fail'' in refinement is
proved, that our TA abstraction will have been adhered to.

Note further that, at seL4's abstract specification level,
we are only able to track addresses this way and enforce
that all accesses conform to that tracking
(through subsequent ``no-fail'' proofs)
for user-controlled kernel objects.
As the kernel's non--user-controlled \emph{global} variables are only assigned addresses
further down in seL4's refinement stack, we exclude these from our reasoning at
the abstract specification level. We plan to include all addresses in the
TA set automatically as they are added to the model by each refinement step, to
force a proper treatment of their cache impacts,
e.g.~by a targeted flush of their addresses from off-core caches.

An important question to ask is: when do we ever remove addresses from the touched addresses set?
The important property this set requires is that no addresses are removed from it \emph{during the
transition being examined}, if we want to reason about that transition's hardware interaction trace.
Therefore it would be safe for the touched addresses set to be emptied \emph{between} or
\emph{immediately at the beginning of} each state transition that is being modeled for time protection purposes
(i.e. it would be fine to empty the set at every kernel entry or exit point in the seL4 kernel model).
In practice, we only empty the touched addresses set in our model during a domain-switch sequence, just
as the cache flushes are performed.
This point is in the ``middle'' of a transition as modeled by seL4's view of the domain-switch transition,
but between two sub-transitions as modeled by the extension.

\subsection{Other impacts on the proof base}\label{s:proof-impacts}
Even after careful planning to minimise its impact, adding touched-addresses tracking to seL4's abstract specification
required essential utility functions and state accessors to be modified, creating a significant amount of
work in returning the proof base to a fully-proved state.
Here we will enumerate the various parts of the proof base that are affected, our plan to make these repairs, and how
far along we are at time of writing.

Repairing the \textbf{abstract invariant / functional correctness} proof base is the most significant piece of engineering
required after introducing TA tracking to seL4.
This is a very large proof artifact, and includes hundreds of lemmas
that depend on executions not to
make any changes to the state, a property which is broken at every point we add to the touched addresses set.
To assist in repairing these proofs, we developed systems that identify various functions that do not make any changes
to the state \emph{except for} changes to the touched addresses set, as well as identifying properties/invariants
that are not concerned with the contents of the touched addresses set (which we name \emph{TA agnostic} properties), and
structures that will discharge proofs about a TA-agnostic property being upheld by a function that changes nothing but
the touched addresses set.

The next most difficult repair to the proof base will be in the \textbf{refinement} proofs.
Fixing this proof base will require showing that each memory access in seL4's abstract specification
accesses an address that was previously added to the TA set.
At time of writing, we have not made significant progress on repairing these proofs.

Another property required by the external model is that the TA set be equivalent when observed
by the domain that is executing. This requires some modifications to the unwinding relation in the \textbf{information
flow} proof module, as well as updating associated proofs.

Finally,
the external time protection model requires that the host system's scheduler meet certain requirements.
Aligning these with seL4's domain scheduler
will be straightforward with the addition of these requirements to
the abstract specification, e.g.~that each domain be assigned adequate time to leave room for time protection
WCET padding operations.

\subsection{Impacts on the kernel design}\label{s:kernel-design}
Our port of seL4's time protection mechanisms to RISC-V also differs from their
initial, unverified implementations for ARM and x86 \citep{Ge_YCH_19} to make
them more amenable to verification; some of these code simplifications were
found to be possible through our formal reasoning.
Our empirical results show that these changes make no difference to their
effectiveness (see \autoref{s:benchmarks}).

For instance, we reasoned we can invoke microarchitectural flushes earlier
on during the domain-switch process, instead of deferring them until just
before returning to the user, as a code audit shows that all addresses touched
after that point belong to the new domain, and thus make no difference to
the provability of the time protection property.

As for changes that simplify verification, we have reduced the number of
kernel API changes needed by initialising kernel image clones at boot time
according to a static configuration, instead of making them the responsibility
of the initial user thread as was done by \citet{Ge_YCH_19}.
Finally, we found that while \citeauthor{Ge_YCH_19} copied the kernel
stack between images on each domain switch, this is redundant as the
stack content is deterministic at this point. This obviates the need for reasoning about the
``dirty'' sub-transition of the time protection mechanism sequence
mentioned in \autoref{s:enforcing}.

\subsection{Directions for refinement}\label{s:refinement-directions}
We now turn to the question of proving time protection for
seL4's C and binary semantics,
which are both related to its abstract specification by refinement
composed of two successive steps (abstract to C \citep{Klein_EHACDEEKNSTW_09},
then C to binary \citep{Sewell_MK_13}).

In these refinement steps we will discover whether we predicted all touched memory
locations correctly. For refinement to hold, the TA set of a lower level
must be a subset of those predicted in the abstract specification. If during
the proof we find that they are not, they will have to be added in the abstract
specification and the information-flow theorem will have to be re-proved.
Furthermore, domain switch sequences will be composed of invocations of
the machine specification, which remains the same between
all three levels.
Therefore we expect to prove the following:

\begin{proposition} [seL4 refinement preserves partition subset invariant and domain switch sequences]
\label{thm:refinement-preserves-reqs}
Since refinement preserves all invariants, it trivially also preserves the
partition subset invariant wherever it originally held. It
leaves domain switch sequences unmodified.
\end{proposition}

Note, that seL4's semantics, that is proved preserved by refinement between all three levels,
excludes the precise memory interaction order.
This remains appropriate for the C semantics because the C compiler
used to compile it to binary
is always liable to introduce reorderings of accesses to memory.
\renewcommand{\theoremautorefname}{Proposition}
Nevertheless, as \autoref{thm:refinement-preserves-reqs} would give us that refinement
\renewcommand{\theoremautorefname}{Theorem}
also preserves the information flow statement, we expect to obtain the best time protection
argument we can hope to apply soundly at the C level:
\begin{proposition} [seL4 refinement to C preserves time protection]
If $\confu$ holds for the seL4 abstract specification
then $\confu_{\mu C}$ holds for the time-protection extension
of its C semantics.
\end{proposition}
Here, $\confu_{\mu C}$ is a time protection property like $\confu_\mu$ but
suitably adapted to the C specification-level state.

\section{Empirical Measurements}\label{s:benchmarks}

Our work uses the FPGA-based CVA-6 core with the \code{fence.t}
instruction of~\citet{Wistoff_SGHB_23}. Consequently, measurements of
channel capacity show the same results, indicating that
all measured channels are closed.

The one significant difference in setup to this earlier work is that
\citeauthor{Wistoff_SGHB_23} uses the dynamically cloned kernel of
\citet{Ge_YCH_19}, while we use a static, two-kernel configuration,
where kernels are switched at the time of the domain switch of the
time-partitioning \emph{domain scheduler} of
\citet{Murray_MBGBSLGK_13}. To demonstrate that this setup does the
job, we repeat the attack the separate kernel images are designed to
prevent: the timing channel through the kernel's code footprint in the
last-level cache. Specifically, the Trojan does nothing to send a
value of \code{0}, or executes a specific system call to send a
\code{1}. The spy measures the latency of executing the same system
call.

\begin{figure}[t]
  \begin{subfigure}[l]{0.45\textwidth}
    \centering
    \includegraphics[width=\textwidth]{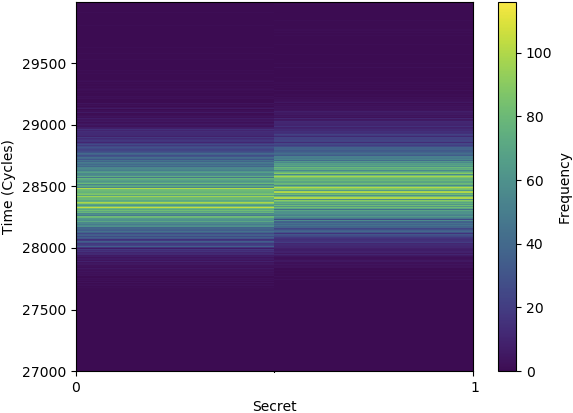}
    \caption{With shared kernel image: \boldmath$M=9.6mb$, $M_0=0.1mb$}
    \label{fig:no-kimage-clones}
  \end{subfigure}
  \hfill
  \begin{subfigure}[l]{0.45\textwidth}
    \centering
    \includegraphics[width=\textwidth]{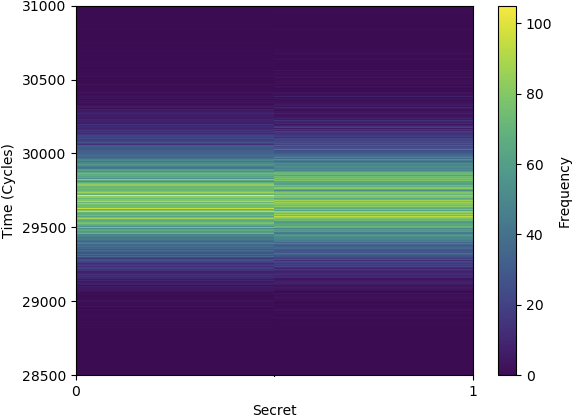}
    \caption{With per-domain kernel images: \boldmath$M=0.1mb$, $M_0=0.1mb$}
    \label{fig:kimage-clones}
  \end{subfigure}
  \caption{Channel matrices for kernel-text channel (100,000 samples).}
  \label{fig:kimage-channel-benchmarks}
\end{figure}

\autoref{fig:no-kimage-clones} shows the channel matrix (distribution
of latencies measured by the spy as a function of the data sent by the
Trojan) if both domains share a kernel image. The distributions
clearly differ, and we measure a small but
significant channel capacity of \(M=9.6\)\,millibits (mb). The 95\%
confidence interval of the apparent channel resulting from statistical
sampling error,  \(\mathcal{M}_0\)~\citep{Ge_YCH_19} is \(0.1\)\,mb in
this case, which indicates that the result is statistically
significant.

With separate kernel images, we get \(M=\mathcal{M}_0=0.1\)\,mb,
indicating the channel is closed. The channel matrix, shown in
\ref{fig:kimage-clones}, shows random fluctuations around the same mean.

\section{Discussion / Lessons Learned}\label{s:discussion}

\subsection{On the effectiveness of determinising the cachedness of kernel global data}\label{s:prefetch}
\citet{Ge_YCH_19} identified a channel via kernel global data.
Some kernel global data cannot be cloned, and its footprint in the off-core
caches therefore presented the opportunity for an inter-domain timing channel.
The authors claimed to patch this channel by determinising the cachedness
of these addresses through \emph{prefetching} -- that is,
by accessing all of them sequentially during a domain-switch.

While examining the details of this countermeasure for formal verification,
we discovered that such an approach does not, in fact, put the cache state
belonging to this set of addresses into a deterministic state.
In fact, without having reliable formal semantics for the specific cache
eviction strategies implemented in particular hardware (generally
unavailable for modern hardware) we concluded that no set of normal
memory interactions can reliably determinise the cache state for any
address or set of addresses.

\citet{Sison_BMKH_23} claimed that such a channel could only be mitigated
by a hardware primitive that meets the explicit requirement of determinising
such state. We have formalised this in our time protection models, and
propose that an off-core cache flushing primitive by address or
cache set that reliably resets any additional state influencing the cache's
replacement policy would be sufficient.

However,
on more detailed investigation
we now believe that that shared kernel globals can be reduced to deterministically accessed public data --
e.g.~domain scheduling data, which can be accessed by no syscalls and is
only accessed during the domain-switch sequence and always in the same order.
We believe this would obviate the need for determinising the cachedness of that
data using either previous methods (prefetch) or a targeted off-core flush;
however, we have the formal means of reasoning about such measures if future versions of the kernel render them necessary.

\subsection{On extending this model to consider other kinds of microarchitectural state}
From our experience in successfully reducing the capture of
microarchitectural state changes to the collection of
the TA set, we learned that we only need to extract interaction data
from the host model at a granularity matching that of the channel elimination strategy:
we do not require any knowledge about accesses to on-core caches because they are always flushed anyway, and
we only need to know about which partition is accessed by each transition.

We see this as a lesson we can apply in future if we must extend our model to
any other stateful microarchitectural feature that is not interacted with via memory
accesses, but still can affect the timing of executions.
Such a feature would not be
represented in our existing microarchitectural model,
as the TA set would contain
no information about accesses to this feature.
However, the approach we have taken thus far in modeling features would make such an adaptation
relatively painless;
again, we would take the approach of only extracting just enough interaction data from the host model (in this case the seL4
abstract specification) at a granularity to match that of the protection mechanism:
\begin{itemize}
\item
A feature whose state is fully flushed on every domain-switch could be considered part of the
\emph{flushable state}, and no interaction details are needed; the model could
quantify over all possible modifications to the feature's state, and as long as a flush operation
is available and called alongside existing flushing mechanisms, our model will show that no
information can be transmitted via that channel.
\item
If the feature's state could be partitioned, it would be considered part of the \emph{partitionable state},
and we would only need to extract enough interaction data from the model to know that accesses are
to the appropriate partition.
\end{itemize}

Robust time protection at the OS level will always require some
heavy-handed reset operation, as a surgical bit-by-bit cleansing of microarchitectural state will be intractable.
The upside of this is that formal reasoning about time protection at the operating system level
is also free of bit-by-bit complications, and is satisfied with input data at the same coarse
granularity as the elimination strategy.

\subsection{On how to handle cross-domain Notifications}
So far we have only described a system configured as a separation kernel,
where no two domains have any allowed communication channels between them.
The next step is to model, and show the absence of timing channels in,
a system where some domains can communicate with each other through
explicitly allowed channels.
These explicit channels are not necessarily bidirectional; some domain $A$
may be able to send messages to $B$, but the policy does not allow any information
to flow from $B$ to $A$ (i.e.~$A \leadsto B$ but $B \not\leadsto A$).
An example of this would be Notifications --- a fast and simple
one-way cross-domain communication in seL4.

In existing seL4 implementations, if user $A$ sends a notification to user $B$,
this will involve the kernel writing to parts of $B$'s memory, an operation
that might be faster or slower depending on the state of $B$'s partitioned
microarchitecture.
Thus, the time taken for $A$ to \emph{send} a notification will yield secrets
about $B$, causing a \emph{backflow} from $B$ to $A$
that we must eliminate if the information-flow policy disallows it.

Further to this off-core timing channel, there are secrets in $B$ that will affect
which memory addresses will be accessed while ``writing'' this notification,
and thereby impact on the non-partitioned on-core cache.
Therefore, simply padding the ``send'' time to some WCET will not eliminate the channel,
as the on-core cache will reveal secrets about $B$'s readiness to receive a message.

To facilitate the extra channel elimination strategies needed,
we anticipate that such cross-domain notifications will need to be added as a
separate kernel API to seL4's existing notifications to avoid
negatively impacting the latter's performance.
Moreover, we expect their verification approach to involve development of a more
sophisticated variant of the ``dirty'' transition that is currently an
optional step in our time protection sequence model of \autoref{s:enforcing}.

\section{Related work}\label{s:related}

\citet{Ge_YH_18} analysed hardware mechanisms for preventing
\utcs and found that there was insufficient support on contemporary
x86 and Arm hardware, leading to a call for an improved
hardware-software contract, specifically the requirement for all
non-partitionable state to be flushable. \citet{Wistoff_SGHB_23}
propose the \code{fence.t} instruction which performs a flush of all
on-core microarchitectural state, by resetting all registers that are
not explicitly part of architected state. \citet{Bourgeat2019MI6}
propose a \texttt{purge} instruction that flushes on-core
microarchitectural components to secure enclaves, a similar proposal
is the \code{FenceX} instruction proposed by~\citet{Li2020SIMF}.

\citet{Escouteloup_LFL_21} propose an alternative design that
introduces \emph{Dome~IDs}, representing security
domains. Microarchitectural state is implicitly flushed when changing
the current Dome~ID. This proposal is presently under discussion in
the RISC-V Microarchitecture Side Channels Special Interest
Group~\citep{uSC-sig}. Our approach, presently based on
\code{fence.t}, can readily adapt to such an alternative.

Several proposals have been made to partition L1
caches~\citep{Percival_05, Page2005Partition, Domnitser_JLAP_12, Dessouky2021ChunkedCache}, while
\citet{Wang_Lee_07, Wang_Lee_08} proposed support for locking cache lines. Such
approaches do not scale to all microarchitectural channels, except in the
extreme case of replicating all state, which would essentially
replicate the core.

The use of multiple kernel images has in the past been proposed for
improving scalability on multicore
platforms~\citep{Boyd-Wickizer_CCMKMPSWDZZ_08, Nightingale_HMHH_09,
  Baumann_BDHIPRSS_09} or hot-plugging of CPU
cores~\citep{Zellweger_GKR_14}. \citet{Ge_YCH_19} proposed per-domain
kernel images for preventing cache channels through shared kernel text
or data. While their model provides dynamic cloning of a kernel, we
replicate at system initialisation time, and statically associate a separate
kernel image with each security domain of seL4's \emph{domain
  scheduler} configuration~\citep{Murray_MBGBSLGK_13}.

Prior proofs of OS-enforced confidentiality~\citep{Murray_MBGBSLGK_13,Costanzo_SG_16, Li_LGNH_21}
only deal with storage channels through architected state.
\citet{Barthe_BCL_12} prove elimination of cache-based
timing channels through a full cache flush; this approach is
prohibitively expensive~\citep{Ge_YCH_19}. \code{stealthMem},
which reserves some cache colours for safe cryptographic operations,
was verified to prevent side-channel leakage \citep{Kim_PM_12,
  Barthe_BCLP_14} but cannot protect against a Trojan
accessing other (non-``stealth'') memory areas.
To our knowledge, the only prior formalisation of OS security to deal
simultaneously with both partitionable and flushable state, as required by
time protection, is the prior formalisation of time protection by
\citet{Sison_BMKH_23}, on which ours is based.

Finally, \citet{Liu_RSGCKY_19} prove a notion of temporal integrity they call
\emph{temporal isolation} in a real-time extension of mCertiKOS, but their
approach concerns the elimination of storage channels via scheduling state and
does not attempt to prevent microarchitectural timing channels.

\section{Conclusions}\label{s:conclusion}

% broad problem for OSes

We have reduced the unknown problem of time protection enforcement by OS kernels to
the more familiar verification problem of collecting a set of touched addresses,
proving a small number of new invariants on that set
and preserving them over refinement.

To this end, we contribute new principles to enable proofs of time-protection properties
and their refinement at different levels of abstraction of an OS kernel specification.

We provide an approach for extending existing OS security models
to account adequately for time protection.
The key conceptual contribution here is the tracking of touched addresses and the use of the
extension of that tracking in a way that solves the unknown-trace problem.

For these extensions we provide conditional proof that, given a proof of the
the original model's security with the tracking added, yields a proof that its
extension satisfies a security property that enforces time protection.

This applies to any OS kernel that satisfies the requirements.

% sel4 specifically

While this proof depends on a conditional statement, there is the following
evidence that the assurance of time protection by the seL4 kernel has already
increased:
\begin{enumerate}
\item
We provide a complete specification of what is necessary to prove it.
\item
We have made changes to the kernel implementation based on this specification.
\item
Our measurements agree with formal predictions.
\end{enumerate}

In doing so, we clarify the required hardware--software contract with respect to
previous formalisations of time protection.

\FIXME{If accepted, re-enable URLs, DOIs!}

\begin{acks}
We thank Thomas Bove for his help in running benchmarks.
Affiliations listed for each author are as at the time their work was carried
out; contact email addresses listed are current at the time of this revision.
This paper describes research that was co-funded by
the Australian Research Council (ARC Project ID DP190103743).
\end{acks}

\balance
{\sloppy
  \bibliographystyle{ACM-Reference-Format}
  \ifPreprint % use the bbl directly
  \bibliography{paper}
  \else
  \bibliography{references}


\providecommand{\noopsort}[1]{}\providecommand{\url}{\error{The bib files now
  require the `url' package!}}\providecommand{\NoRemove}{}
\begin{thebibliography}{60}

%%% the Web address.

\ifx \showCODEN    \undefined \def \showCODEN     #1{\unskip}     \fi
\ifx \showDOI      \undefined \def \showDOI       #1{#1}\fi
\ifx \showISBNx    \undefined \def \showISBNx     #1{\unskip}     \fi
\ifx \showISBNxiii \undefined \def \showISBNxiii  #1{\unskip}     \fi
\ifx \showISSN     \undefined \def \showISSN      #1{\unskip}     \fi
\ifx \showLCCN     \undefined \def \showLCCN      #1{\unskip}     \fi
\ifx \shownote     \undefined \def \shownote      #1{#1}          \fi
\ifx \showarticletitle \undefined \def \showarticletitle #1{#1}   \fi
\ifx \showURL      \undefined \def \showURL       {\relax}        \fi
% invisible to TeX
\providecommand\bibfield[2]{#2}
\providecommand\bibinfo[2]{#2}
\providecommand\natexlab[1]{#1}
\providecommand\showeprint[2][]{arXiv:#2}

\bibitem[\protect\citeauthoryear{Backes, D\"{u}rmuth, Gerling, Pinkal, and
  Sporleder}{Backes et~al\mbox{.}}{2010}]%
        {Backes_DGPS_10}
\bibfield{author}{\bibinfo{person}{Michael Backes}, \bibinfo{person}{Markus
  D\"{u}rmuth}, \bibinfo{person}{Sebastian Gerling}, \bibinfo{person}{Manfred
  Pinkal}, {and} \bibinfo{person}{Caroline Sporleder}.}
  \bibinfo{year}{2010}\natexlab{}.
\newblock \showarticletitle{Acoustic side-channel attacks on printers}. In
  \bibinfo{booktitle}{\emph{Proceedings of the 19th USENIX Security
  Symposium}}. \bibinfo{publisher}{USENIX}, \bibinfo{address}{Washington, DC},
  \bibinfo{pages}{1--16}.
\newblock


\bibitem[\protect\citeauthoryear{Barthe, Betarte, Campo, and Luna}{Barthe
  et~al\mbox{.}}{2012}]%
        {Barthe_BCL_12}
\bibfield{author}{\bibinfo{person}{Gilles Barthe}, \bibinfo{person}{Gustavo
  Betarte}, \bibinfo{person}{Juan~Diego Campo}, {and} \bibinfo{person}{Carlos
  Luna}.} \bibinfo{year}{2012}\natexlab{}.
\newblock \showarticletitle{Cache-Leakage Resilient {OS} Isolation in an
  Idealized Model of Virtualization}. In \bibinfo{booktitle}{\emph{Proceedings
  of the 25th IEEE Computer Security Foundations Symposium}}.
  \bibinfo{publisher}{IEEE}, \bibinfo{pages}{186--197}.
\newblock
\urldef\tempurl%
\url{https://doi.org/10.1109/CSF.2012.17}
\showDOI{\tempurl}


\bibitem[\protect\citeauthoryear{Barthe, Betarte, Campo, Luna, and
  Pichardie}{Barthe et~al\mbox{.}}{2014}]%
        {Barthe_BCLP_14}
\bibfield{author}{\bibinfo{person}{Gilles Barthe}, \bibinfo{person}{Gustavo
  Betarte}, \bibinfo{person}{Juan~Diego Campo}, \bibinfo{person}{Carlos~Daniel
  Luna}, {and} \bibinfo{person}{David Pichardie}.}
  \bibinfo{year}{2014}\natexlab{}.
\newblock \showarticletitle{System-level Non-interference for Constant-time
  Cryptography}. In \bibinfo{booktitle}{\emph{Proceedings of the 2014 {ACM}
  {SIGSAC} Conference on Computer and Communications Security, Scottsdale, AZ,
  USA, November 3-7, 2014}}. \bibinfo{publisher}{{ACM}},
  \bibinfo{pages}{1267--1279}.
\newblock
\urldef\tempurl%
\url{https://doi.org/10.1145/2660267.2660283}
\showDOI{\tempurl}


\bibitem[\protect\citeauthoryear{Baumann, Barham, Dagand, Harris, Isaacs,
  Peter, Roscoe, Sch{\"u}pbach, and Singhania}{Baumann et~al\mbox{.}}{2009}]%
        {Baumann_BDHIPRSS_09}
\bibfield{author}{\bibinfo{person}{Andrew Baumann}, \bibinfo{person}{Paul
  Barham}, \bibinfo{person}{Pierre-Evariste Dagand}, \bibinfo{person}{Tim
  Harris}, \bibinfo{person}{Rebecca Isaacs}, \bibinfo{person}{Simon Peter},
  \bibinfo{person}{Timothy Roscoe}, \bibinfo{person}{Adrian Sch{\"u}pbach},
  {and} \bibinfo{person}{Akhilesh Singhania}.} \bibinfo{year}{2009}\natexlab{}.
\newblock \showarticletitle{The Multikernel: A New {OS} Architecture for
  Scalable Multicore Systems}. In \bibinfo{booktitle}{\emph{ACM Symposium on
  Operating Systems Principles}}. \bibinfo{publisher}{ACM},
  \bibinfo{address}{Big Sky, MT, US}, \bibinfo{pages}{29--44}.
\newblock


\bibitem[\protect\citeauthoryear{Bourgeat, Lebedev, Wright, Zhang, Arvind, and
  Devadas}{Bourgeat et~al\mbox{.}}{2019}]%
        {Bourgeat2019MI6}
\bibfield{author}{\bibinfo{person}{Thomas Bourgeat}, \bibinfo{person}{Ilia~A.
  Lebedev}, \bibinfo{person}{Andrew Wright}, \bibinfo{person}{Sizhuo Zhang},
  \bibinfo{person}{Arvind}, {and} \bibinfo{person}{Srinivas Devadas}.}
  \bibinfo{year}{2019}\natexlab{}.
\newblock \showarticletitle{{MI6}: Secure Enclaves in a Speculative
  Out-of-Order Processor}.
\newblock \bibinfo{journal}{\emph{Proceedings of the 52nd Annual IEEE/ACM
  International Symposium on Microarchitecture (MICRO)}}
  (\bibinfo{year}{2019}), \bibinfo{pages}{42--56}.
\newblock
\urldef\tempurl%
\url{https://doi.org/10.1145/3352460.3358310}
\showDOI{\tempurl}


\bibitem[\protect\citeauthoryear{Boyd-Wickizer, Chen, Chen, Mao, Kaashoek,
  Morris, Pesterev, Stein, Wu, Dai, Zhang, and Zhang}{Boyd-Wickizer
  et~al\mbox{.}}{2008}]%
        {Boyd-Wickizer_CCMKMPSWDZZ_08}
\bibfield{author}{\bibinfo{person}{Silas Boyd-Wickizer}, \bibinfo{person}{Haibo
  Chen}, \bibinfo{person}{Rong Chen}, \bibinfo{person}{Yandong Mao},
  \bibinfo{person}{Frans Kaashoek}, \bibinfo{person}{Robert Morris},
  \bibinfo{person}{Aleksey Pesterev}, \bibinfo{person}{Lex Stein},
  \bibinfo{person}{Ming Wu}, \bibinfo{person}{Yuehua Dai},
  \bibinfo{person}{Yang Zhang}, {and} \bibinfo{person}{Zheng Zhang}.}
  \bibinfo{year}{2008}\natexlab{}.
\newblock \showarticletitle{{Corey}: an operating system for many cores}. In
  \bibinfo{booktitle}{\emph{Proceedings of the 8th USENIX Symposium on
  Operating Systems Design and Implementation}}. \bibinfo{publisher}{USENIX},
  \bibinfo{address}{San Diego, CA, US}, \bibinfo{pages}{43--57}.
\newblock


\bibitem[\protect\citeauthoryear{Costanzo, Shao, and Gu}{Costanzo
  et~al\mbox{.}}{2016}]%
        {Costanzo_SG_16}
\bibfield{author}{\bibinfo{person}{David Costanzo}, \bibinfo{person}{Zhong
  Shao}, {and} \bibinfo{person}{Ronghui Gu}.} \bibinfo{year}{2016}\natexlab{}.
\newblock \showarticletitle{End-to-end verification of information-flow
  security for {C} and assembly programs}. In \bibinfo{booktitle}{\emph{ACM
  SIGPLAN Conference on Programming Language Design and Implementation}}.
  \bibinfo{pages}{648--664}.
\newblock


\bibitem[\protect\citeauthoryear{de~Roever and Engelhardt}{de~Roever and
  Engelhardt}{1998}]%
        {deRoever_Engelhardt:DR}
\bibfield{author}{\bibinfo{person}{Willem-Paul de Roever} {and}
  \bibinfo{person}{Kai Engelhardt}.} \bibinfo{year}{1998}\natexlab{}.
\newblock \bibinfo{booktitle}{\emph{Data Refinement: Model-Oriented Proof
  Methods and their Comparison}}.
\newblock Number~47 in \bibinfo{series}{Cambridge Tracts in Theoretical
  Computer Science}. \bibinfo{publisher}{Cambridge University Press},
  \bibinfo{address}{United Kingdom}.
\newblock


\bibitem[\protect\citeauthoryear{Department of Defence}{Department of
  Defence}{1986}]%
        {DoD_85:orange}
Department of Defence \bibinfo{year}{1986}\natexlab{}.
\newblock \bibinfo{booktitle}{\emph{Trusted Computer System Evaluation
  Criteria}}.
\newblock Department of Defence.
\newblock
\newblock
\shownote{{DoD 5200.28-STD}.}


\bibitem[\protect\citeauthoryear{Dessouky, Gruler, Mahmoody, Sadeghi, and
  Stapf}{Dessouky et~al\mbox{.}}{2021}]%
        {Dessouky2021ChunkedCache}
\bibfield{author}{\bibinfo{person}{Ghada Dessouky}, \bibinfo{person}{Alexander
  Gruler}, \bibinfo{person}{Pouya Mahmoody}, \bibinfo{person}{Ahmad-Reza
  Sadeghi}, {and} \bibinfo{person}{Emmanuel Stapf}.}
  \bibinfo{year}{2021}\natexlab{}.
\newblock \showarticletitle{Chunked-Cache: On-Demand and Scalable Cache
  Isolation for Security Architectures}.
\newblock \bibinfo{journal}{\emph{ArXiv}}  \bibinfo{volume}{abs/2110.08139}
  (\bibinfo{year}{2021}).
\newblock


\bibitem[\protect\citeauthoryear{Deutsch, Yang, Bourgeat, Drean, Emer, and
  Yan}{Deutsch et~al\mbox{.}}{2022}]%
        {Deutsch_YBDEY_22}
\bibfield{author}{\bibinfo{person}{Peter~W. Deutsch}, \bibinfo{person}{Yuheng
  Yang}, \bibinfo{person}{Thomas Bourgeat}, \bibinfo{person}{Jules Drean},
  \bibinfo{person}{Joel~S. Emer}, {and} \bibinfo{person}{Mengjia Yan}.}
  \bibinfo{year}{2022}\natexlab{}.
\newblock \showarticletitle{{DAGguise}: mitigating memory timing side
  channels}. In \bibinfo{booktitle}{\emph{International Conference on
  Architectural Support for Programming Languages and Operating Systems}}.
  \bibinfo{pages}{329–--343}.
\newblock
\urldef\tempurl%
\url{https://doi.org/10.1145/3582016.3582033}
\showDOI{\tempurl}


\bibitem[\protect\citeauthoryear{Domnitser, Jaleel, Loew, Abu-Ghazaleh, and
  Ponomarev}{Domnitser et~al\mbox{.}}{2012}]%
        {Domnitser_JLAP_12}
\bibfield{author}{\bibinfo{person}{Leonid Domnitser}, \bibinfo{person}{Aamer
  Jaleel}, \bibinfo{person}{Jason Loew}, \bibinfo{person}{Nael Abu-Ghazaleh},
  {and} \bibinfo{person}{Dmitry Ponomarev}.} \bibinfo{year}{2012}\natexlab{}.
\newblock \showarticletitle{Non-Monopolizable Caches: Low-Complexity Mitigation
  of Cache Side Channel Attacks}.
\newblock \bibinfo{journal}{\emph{ACM Transactions on Architecture and Code
  Optimization}} \bibinfo{volume}{8}, \bibinfo{number}{4} (\bibinfo{date}{Jan.}
  \bibinfo{year}{2012}).
\newblock


\bibitem[\protect\citeauthoryear{Escouteloup, Lashermes, Fournier, and
  Lanet}{Escouteloup et~al\mbox{.}}{2021}]%
        {Escouteloup_LFL_21}
\bibfield{author}{\bibinfo{person}{Mathieu Escouteloup}, \bibinfo{person}{Ronan
  Lashermes}, \bibinfo{person}{Jacques Fournier}, {and}
  \bibinfo{person}{Jean-Louis Lanet}.} \bibinfo{year}{2021}\natexlab{}.
\newblock \showarticletitle{Under the dome: preventing hardware timing
  information leakage}. In \bibinfo{booktitle}{\emph{Smart Card Research and
  Advanced Application Conference (CARDIS)}}. \bibinfo{address}{L\"{u}beck,
  DE}, \bibinfo{pages}{1--20}.
\newblock


\bibitem[\protect\citeauthoryear{Fox and Myreen}{Fox and Myreen}{2010}]%
        {Fox_Myreen_10}
\bibfield{author}{\bibinfo{person}{Anthony Fox} {and} \bibinfo{person}{Magnus
  Myreen}.} \bibinfo{year}{2010}\natexlab{}.
\newblock \showarticletitle{A Trustworthy Monadic Formalization of the {ARMv7}
  Instruction Set Architecture}. In \bibinfo{booktitle}{\emph{Proceedings of
  the 1st International Conference on Interactive Theorem Proving}}
  \emph{(\bibinfo{series}{Lecture Notes in Computer Science})},
  Vol.~\bibinfo{volume}{6172}. \bibinfo{publisher}{Springer},
  \bibinfo{address}{Edinburgh, UK}, \bibinfo{pages}{243--258}.
\newblock


\bibitem[\protect\citeauthoryear{Ge, Yarom, Chothia, and Heiser}{Ge
  et~al\mbox{.}}{2019}]%
        {Ge_YCH_19}
\bibfield{author}{\bibinfo{person}{Qian Ge}, \bibinfo{person}{Yuval Yarom},
  \bibinfo{person}{Tom Chothia}, {and} \bibinfo{person}{Gernot Heiser}.}
  \bibinfo{year}{2019}\natexlab{}.
\newblock \showarticletitle{Time Protection: the Missing {OS} Abstraction}. In
  \bibinfo{booktitle}{\emph{EuroSys Conference}}. \bibinfo{publisher}{ACM},
  \bibinfo{address}{Dresden, Germany}, 17.
\newblock


\bibitem[\protect\citeauthoryear{Ge, Yarom, Cock, and Heiser}{Ge
  et~al\mbox{.}}{2018b}]%
        {Ge_YCH_18}
\bibfield{author}{\bibinfo{person}{Qian Ge}, \bibinfo{person}{Yuval Yarom},
  \bibinfo{person}{David Cock}, {and} \bibinfo{person}{Gernot Heiser}.}
  \bibinfo{year}{2018}\natexlab{b}.
\newblock \showarticletitle{{A} Survey of Microarchitectural Timing Attacks and
  Countermeasures on Contemporary Hardware}.
\newblock \bibinfo{journal}{\emph{Journal of Cryptographic Engineering}}
  \bibinfo{volume}{8} (\bibinfo{date}{April} \bibinfo{year}{2018}),
  \bibinfo{pages}{1--27}.
\newblock
\urldef\tempurl%
\url{https://doi.org/10.1007/s13389-016-0141-6}
\showDOI{\tempurl}


\bibitem[\protect\citeauthoryear{Ge, Yarom, and Heiser}{Ge
  et~al\mbox{.}}{2018a}]%
        {Ge_YH_18}
\bibfield{author}{\bibinfo{person}{Qian Ge}, \bibinfo{person}{Yuval Yarom},
  {and} \bibinfo{person}{Gernot Heiser}.} \bibinfo{year}{2018}\natexlab{a}.
\newblock \showarticletitle{No Security Without Time Protection: We Need a New
  Hardware-Software Contract}. In \bibinfo{booktitle}{\emph{Asia-Pacific
  Workshop on Systems (APSys)}}. \bibinfo{publisher}{ACM SIGOPS},
  \bibinfo{address}{Korea}, 9.
\newblock
\urldef\tempurl%
\url{https://doi.org/10.1145/3265723.3265724}
\showDOI{\tempurl}


\bibitem[\protect\citeauthoryear{Genkin, Pachmanov, Pipman, and Tromer}{Genkin
  et~al\mbox{.}}{2015}]%
        {Genkin_PPT_15}
\bibfield{author}{\bibinfo{person}{Daniel Genkin}, \bibinfo{person}{Lev
  Pachmanov}, \bibinfo{person}{Itamar Pipman}, {and} \bibinfo{person}{Eran
  Tromer}.} \bibinfo{year}{2015}\natexlab{}.
\newblock \showarticletitle{Stealing Keys from {PCs} Using a Radio: Cheap
  Electromagnetic Attacks on Windowed Exponentiation}. In
  \bibinfo{booktitle}{\emph{Workshop on Cryptographic Hardware and Embedded
  Systems}}. \bibinfo{address}{Saint Malo, FR}, \bibinfo{pages}{207--228}.
\newblock


\bibitem[\protect\citeauthoryear{Genkin, Shamir, and Tromer}{Genkin
  et~al\mbox{.}}{2014}]%
        {Genkin_ST_14}
\bibfield{author}{\bibinfo{person}{Daniel Genkin}, \bibinfo{person}{Adi
  Shamir}, {and} \bibinfo{person}{Eran Tromer}.}
  \bibinfo{year}{2014}\natexlab{}.
\newblock \showarticletitle{{RSA} Key Extraction via Low-Bandwidth Acoustic
  Cryptanalysis}. In \bibinfo{booktitle}{\emph{International Cryptology
  Conference---CRYPTO 2014}}. \bibinfo{address}{Santa Barbara, CA, US},
  \bibinfo{pages}{444--461}.
\newblock


\bibitem[\protect\citeauthoryear{Heiser, Klein, and Murray}{Heiser
  et~al\mbox{.}}{2019}]%
        {Heiser_KM_19}
\bibfield{author}{\bibinfo{person}{Gernot Heiser}, \bibinfo{person}{Gerwin
  Klein}, {and} \bibinfo{person}{Toby Murray}.}
  \bibinfo{year}{2019}\natexlab{}.
\newblock \showarticletitle{Can We Prove Time Protection?}. In
  \bibinfo{booktitle}{\emph{Workshop on Hot Topics in Operating Systems
  (HotOS)}}. \bibinfo{publisher}{ACM}, \bibinfo{address}{Bertinoro, Italy},
  \bibinfo{pages}{23--29}.
\newblock
\urldef\tempurl%
\url{https://doi.org/10.1145/3317550.3321431}
\showDOI{\tempurl}


\bibitem[\protect\citeauthoryear{Hu}{Hu}{1991}]%
        {Hu_91}
\bibfield{author}{\bibinfo{person}{Wei-Ming Hu}.}
  \bibinfo{year}{1991}\natexlab{}.
\newblock \showarticletitle{Reducing timing channels with fuzzy time}. In
  \bibinfo{booktitle}{\emph{Proceedings of the 1991 IEEE Computer Society
  Symposium on Research in Security and Privacy}}. \bibinfo{publisher}{IEEE
  Computer Society}, \bibinfo{address}{Oakland, CA, US},
  \bibinfo{pages}{8--20}.
\newblock


\bibitem[\protect\citeauthoryear{Kessler and Hill}{Kessler and Hill}{1992}]%
        {Kessler_Hill_92}
\bibfield{author}{\bibinfo{person}{R.~E. Kessler} {and}
  \bibinfo{person}{Mark~D. Hill}.} \bibinfo{year}{1992}\natexlab{}.
\newblock \showarticletitle{Page placement algorithms for large real-indexed
  caches}.
\newblock \bibinfo{journal}{\emph{ACM Transactions on Computer Systems}}
  \bibinfo{volume}{10} (\bibinfo{year}{1992}), \bibinfo{pages}{338--359}.
\newblock


\bibitem[\protect\citeauthoryear{Kim, Peinado, and Mainar-Ruiz}{Kim
  et~al\mbox{.}}{2012}]%
        {Kim_PM_12}
\bibfield{author}{\bibinfo{person}{Taesoo Kim}, \bibinfo{person}{Marcus
  Peinado}, {and} \bibinfo{person}{Gloria Mainar-Ruiz}.}
  \bibinfo{year}{2012}\natexlab{}.
\newblock \showarticletitle{\textsc{StealthMem}: system-level protection
  against cache-based side channel attacks in the cloud}. In
  \bibinfo{booktitle}{\emph{Proceedings of the 21st USENIX Security
  Symposium}}. \bibinfo{publisher}{USENIX}, \bibinfo{address}{Bellevue, WA,
  US}, \bibinfo{pages}{189--204}.
\newblock


\bibitem[\protect\citeauthoryear{Klein, Andronick, Elphinstone, Murray, Sewell,
  Kolanski, and Heiser}{Klein et~al\mbox{.}}{2014}]%
        {Klein_AEMSKH_14}
\bibfield{author}{\bibinfo{person}{Gerwin Klein}, \bibinfo{person}{June
  Andronick}, \bibinfo{person}{Kevin Elphinstone}, \bibinfo{person}{Toby
  Murray}, \bibinfo{person}{Thomas Sewell}, \bibinfo{person}{Rafal Kolanski},
  {and} \bibinfo{person}{Gernot Heiser}.} \bibinfo{year}{2014}\natexlab{}.
\newblock \showarticletitle{Comprehensive Formal Verification of an {OS}
  Microkernel}.
\newblock \bibinfo{journal}{\emph{ACM Transactions on Computer Systems}}
  \bibinfo{volume}{32}, \bibinfo{number}{1} (\bibinfo{date}{Feb.}
  \bibinfo{year}{2014}), \bibinfo{pages}{2:1--2:70}.
\newblock
\urldef\tempurl%
\url{https://doi.org/10.1145/2560537}
\showDOI{\tempurl}


\bibitem[\protect\citeauthoryear{Klein, Elphinstone, Heiser, Andronick, Cock,
  Derrin, Elkaduwe, Engelhardt, Kolanski, Norrish, Sewell, Tuch, and
  Winwood}{Klein et~al\mbox{.}}{2009}]%
        {Klein_EHACDEEKNSTW_09}
\bibfield{author}{\bibinfo{person}{Gerwin Klein}, \bibinfo{person}{Kevin
  Elphinstone}, \bibinfo{person}{Gernot Heiser}, \bibinfo{person}{June
  Andronick}, \bibinfo{person}{David Cock}, \bibinfo{person}{Philip Derrin},
  \bibinfo{person}{Dhammika Elkaduwe}, \bibinfo{person}{Kai Engelhardt},
  \bibinfo{person}{Rafal Kolanski}, \bibinfo{person}{Michael Norrish},
  \bibinfo{person}{Thomas Sewell}, \bibinfo{person}{Harvey Tuch}, {and}
  \bibinfo{person}{Simon Winwood}.} \bibinfo{year}{2009}\natexlab{}.
\newblock \showarticletitle{{seL4}: Formal Verification of an {OS} Kernel}. In
  \bibinfo{booktitle}{\emph{ACM Symposium on Operating Systems Principles}}.
  \bibinfo{publisher}{ACM}, \bibinfo{address}{Big Sky, MT, USA},
  \bibinfo{pages}{207--220}.
\newblock


\bibitem[\protect\citeauthoryear{Klein, Murray, Gammie, Sewell, and
  Winwood}{Klein et~al\mbox{.}}{2011}]%
        {Klein_MGSW_11}
\bibfield{author}{\bibinfo{person}{Gerwin Klein}, \bibinfo{person}{Toby
  Murray}, \bibinfo{person}{Peter Gammie}, \bibinfo{person}{Thomas Sewell},
  {and} \bibinfo{person}{Simon Winwood}.} \bibinfo{year}{2011}\natexlab{}.
\newblock \showarticletitle{Provable Security: How feasible is it?}. In
  \bibinfo{booktitle}{\emph{Workshop on Hot Topics in Operating Systems
  (HotOS)}}. \bibinfo{publisher}{USENIX}, \bibinfo{address}{Napa, USA},
  \bibinfo{pages}{5}.
\newblock


\bibitem[\protect\citeauthoryear{Kocher, Horn, Fogh, Genkin, Gruss, Haas,
  Haburg, Lipp, Mangard, Prescher, Schwartz, and Yarom}{Kocher
  et~al\mbox{.}}{2019}]%
        {Kocher_HFGGHHLMPSY_19}
\bibfield{author}{\bibinfo{person}{Paul Kocher}, \bibinfo{person}{Jann Horn},
  \bibinfo{person}{Anders Fogh}, \bibinfo{person}{Daniel Genkin},
  \bibinfo{person}{Daniel Gruss}, \bibinfo{person}{Werner Haas},
  \bibinfo{person}{Mike Haburg}, \bibinfo{person}{Moritz Lipp},
  \bibinfo{person}{Stefan Mangard}, \bibinfo{person}{Thomas Prescher},
  \bibinfo{person}{Michael Schwartz}, {and} \bibinfo{person}{Yuval Yarom}.}
  \bibinfo{year}{2019}\natexlab{}.
\newblock \showarticletitle{Spectre Attacks: Exploiting Speculative Execution}.
  In \bibinfo{booktitle}{\emph{IEEE Symposium on Security and Privacy}}.
  \bibinfo{address}{San Francisco, CA, US}, \bibinfo{pages}{19--37}.
\newblock


\bibitem[\protect\citeauthoryear{Kocher, Jaffe, and Jun}{Kocher
  et~al\mbox{.}}{1999}]%
        {Kocher_JJ_99}
\bibfield{author}{\bibinfo{person}{Paul Kocher}, \bibinfo{person}{Joshua
  Jaffe}, {and} \bibinfo{person}{Benjamin Jun}.}
  \bibinfo{year}{1999}\natexlab{}.
\newblock \showarticletitle{Differential Power Analysis}. In
  \bibinfo{booktitle}{\emph{International Cryptology Conference---CRYPTO}}
  \emph{(\bibinfo{series}{Lecture Notes in Computer Science})},
  Vol.~\bibinfo{volume}{1666}. \bibinfo{publisher}{Springer},
  \bibinfo{pages}{388--397}.
\newblock
\urldef\tempurl%
\url{https://doi.org/10.1007/3-540-48405-1_25}
\showDOI{\tempurl}


\bibitem[\protect\citeauthoryear{Lampson}{Lampson}{1973}]%
        {Lampson_73}
\bibfield{author}{\bibinfo{person}{Butler~W. Lampson}.}
  \bibinfo{year}{1973}\natexlab{}.
\newblock \showarticletitle{A Note on the Confinement Problem}.
\newblock \bibinfo{journal}{\emph{Commun. ACM}}  \bibinfo{volume}{16}
  (\bibinfo{year}{1973}), \bibinfo{pages}{613--615}.
\newblock
\urldef\tempurl%
\url{https://doi.org/10.1145/362375.362389}
\showDOI{\tempurl}


\bibitem[\protect\citeauthoryear{Li, Li, Gu, Nieh, and Hui}{Li
  et~al\mbox{.}}{2021}]%
        {Li_LGNH_21}
\bibfield{author}{\bibinfo{person}{Shih-Wei Li}, \bibinfo{person}{Xupeng Li},
  \bibinfo{person}{Ronghui Gu}, \bibinfo{person}{Jason Nieh}, {and}
  \bibinfo{person}{John~Zhuang Hui}.} \bibinfo{year}{2021}\natexlab{}.
\newblock \showarticletitle{A Secure and Formally Verified {Linux} {KVM}
  Hypervisor}. In \bibinfo{booktitle}{\emph{IEEE Symposium on Security and
  Privacy}}.
\newblock


\bibitem[\protect\citeauthoryear{Li, Hopkins, and Parameswaran}{Li
  et~al\mbox{.}}{2020}]%
        {Li2020SIMF}
\bibfield{author}{\bibinfo{person}{Tuo Li}, \bibinfo{person}{Bradley~D.
  Hopkins}, {and} \bibinfo{person}{Sri Parameswaran}.}
  \bibinfo{year}{2020}\natexlab{}.
\newblock \showarticletitle{SIMF: Single-Instruction Multiple-Flush Mechanism
  for Processor Temporal Isolation}.
\newblock \bibinfo{journal}{\emph{ArXiv}}  \bibinfo{volume}{abs/2011.10249}
  (\bibinfo{year}{2020}).
\newblock


\bibitem[\protect\citeauthoryear{Liedtke, Islam, and Jaeger}{Liedtke
  et~al\mbox{.}}{1997}]%
        {Liedtke_IJ_97}
\bibfield{author}{\bibinfo{person}{Jochen Liedtke}, \bibinfo{person}{Nayeem
  Islam}, {and} \bibinfo{person}{Trent Jaeger}.}
  \bibinfo{year}{1997}\natexlab{}.
\newblock \showarticletitle{Preventing Denial-of-Service Attacks on a
  {$\mu$}-Kernel for {WebOS}es}. In \bibinfo{booktitle}{\emph{Proceedings of
  the 6th Workshop on Hot Topics in Operating Systems (HotOS)}}. IEEE,
  \bibinfo{address}{Cape Cod, MA, US}, \bibinfo{pages}{73--79}.
\newblock


\bibitem[\protect\citeauthoryear{Liu, Rieg, Shao, Gu, Costanzo, Kim, and
  Yoon}{Liu et~al\mbox{.}}{2019}]%
        {Liu_RSGCKY_19}
\bibfield{author}{\bibinfo{person}{Mengqi Liu}, \bibinfo{person}{Lionel Rieg},
  \bibinfo{person}{Zhong Shao}, \bibinfo{person}{Ronghui Gu},
  \bibinfo{person}{David Costanzo}, \bibinfo{person}{Jung-Eun Kim}, {and}
  \bibinfo{person}{Man-Ki Yoon}.} \bibinfo{year}{2019}\natexlab{}.
\newblock \showarticletitle{Virtual timeline: a formal abstraction for
  verifying preemptive schedulers with temporal isolation}.
\newblock \bibinfo{journal}{\emph{Proceedings of the ACM on Programming
  Languages}} \bibinfo{volume}{4}, \bibinfo{number}{POPL}
  (\bibinfo{year}{2019}), \bibinfo{pages}{1--31}.
\newblock


\bibitem[\protect\citeauthoryear{Lynch, Bray, and Flynn}{Lynch
  et~al\mbox{.}}{1992}]%
        {Lynch_BF_92}
\bibfield{author}{\bibinfo{person}{William~L. Lynch}, \bibinfo{person}{Brian~K.
  Bray}, {and} \bibinfo{person}{M.~J. Flynn}.} \bibinfo{year}{1992}\natexlab{}.
\newblock \showarticletitle{The effect of page allocation on caches}. In
  \bibinfo{booktitle}{\emph{ACM/IEE International Symposium on
  Microarchitecture}}. \bibinfo{publisher}{IEEE}, \bibinfo{address}{Portland,
  OR, US}, \bibinfo{pages}{222--225}.
\newblock


\bibitem[\protect\citeauthoryear{Masti, Rai, Ranganathan, M{\"u}ller, Thiele,
  and \v{C}apkun}{Masti et~al\mbox{.}}{2015}]%
        {Masti_RRMTC_15}
\bibfield{author}{\bibinfo{person}{Ramya~Jayaram Masti},
  \bibinfo{person}{Devendra Rai}, \bibinfo{person}{Aanjhan Ranganathan},
  \bibinfo{person}{Christian M{\"u}ller}, \bibinfo{person}{Lothar Thiele},
  {and} \bibinfo{person}{Srdjan \v{C}apkun}.} \bibinfo{year}{2015}\natexlab{}.
\newblock \showarticletitle{Thermal Covert Channels on Multi-core Platforms}.
  In \bibinfo{booktitle}{\emph{Proceedings of the 24th USENIX Security
  Symposium}}. \bibinfo{address}{Washington, DC, US},
  \bibinfo{pages}{865--880}.
\newblock


\bibitem[\protect\citeauthoryear{Molnar, Piotrowski, Schultz, and
  Wagner}{Molnar et~al\mbox{.}}{2006}]%
        {Molnar_PSW_06}
\bibfield{author}{\bibinfo{person}{David Molnar}, \bibinfo{person}{Matt
  Piotrowski}, \bibinfo{person}{David Schultz}, {and} \bibinfo{person}{David
  Wagner}.} \bibinfo{year}{2006}\natexlab{}.
\newblock \showarticletitle{The program counter security model: Automatic
  detection and removal of control-flow side channel attacks}. In
  \bibinfo{booktitle}{\emph{Information Security and Cryptology-ICISC 2005: 8th
  International Conference, Seoul, Korea, December 1-2, 2005, Revised Selected
  Papers 8}}. Springer, \bibinfo{pages}{156--168}.
\newblock


\bibitem[\protect\citeauthoryear{Murdoch}{Murdoch}{2006}]%
        {Murdoch_06}
\bibfield{author}{\bibinfo{person}{Steven~J. Murdoch}.}
  \bibinfo{year}{2006}\natexlab{}.
\newblock \showarticletitle{Hot or not: revealing hidden services by their
  clock skew}. In \bibinfo{booktitle}{\emph{Proceedings of the 13th ACM
  Conference on Computer and Communications Security}}.
  \bibinfo{publisher}{ACM}, \bibinfo{address}{Alexandria, VA, US},
  \bibinfo{pages}{27--36}.
\newblock
\urldef\tempurl%
\url{https://doi.org/10.1145/1180405.1180410}
\showDOI{\tempurl}


\bibitem[\protect\citeauthoryear{Murray, Matichuk, Brassil, Gammie, Bourke,
  Seefried, Lewis, Gao, and Klein}{Murray et~al\mbox{.}}{2013}]%
        {Murray_MBGBSLGK_13}
\bibfield{author}{\bibinfo{person}{Toby Murray}, \bibinfo{person}{Daniel
  Matichuk}, \bibinfo{person}{Matthew Brassil}, \bibinfo{person}{Peter Gammie},
  \bibinfo{person}{Timothy Bourke}, \bibinfo{person}{Sean Seefried},
  \bibinfo{person}{Corey Lewis}, \bibinfo{person}{Xin Gao}, {and}
  \bibinfo{person}{Gerwin Klein}.} \bibinfo{year}{2013}\natexlab{}.
\newblock \showarticletitle{{seL4}: from General Purpose to a Proof of
  Information Flow Enforcement}. In \bibinfo{booktitle}{\emph{IEEE Symposium on
  Security and Privacy}}. \bibinfo{publisher}{IEEE}, \bibinfo{address}{San
  Francisco, CA}, \bibinfo{pages}{415--429}.
\newblock
\urldef\tempurl%
\url{https://doi.org/10.1109/SP.2013.35}
\showDOI{\tempurl}


\bibitem[\protect\citeauthoryear{Murray, Matichuk, Brassil, Gammie, and
  Klein}{Murray et~al\mbox{.}}{2012}]%
        {Murray_MBGK_12}
\bibfield{author}{\bibinfo{person}{Toby Murray}, \bibinfo{person}{Daniel
  Matichuk}, \bibinfo{person}{Matthew Brassil}, \bibinfo{person}{Peter Gammie},
  {and} \bibinfo{person}{Gerwin Klein}.} \bibinfo{year}{2012}\natexlab{}.
\newblock \showarticletitle{Noninterference for Operating System Kernels}. In
  \bibinfo{booktitle}{\emph{International Conference on Certified Programs and
  Proofs}}. \bibinfo{publisher}{Springer}, \bibinfo{address}{Kyoto, Japan},
  \bibinfo{pages}{126--142}.
\newblock


\bibitem[\protect\citeauthoryear{Nightingale, Hodson, McIlroy, Hawblitzel, and
  Hunt}{Nightingale et~al\mbox{.}}{2009}]%
        {Nightingale_HMHH_09}
\bibfield{author}{\bibinfo{person}{Edmund~B. Nightingale},
  \bibinfo{person}{Orion Hodson}, \bibinfo{person}{Ross McIlroy},
  \bibinfo{person}{Chris Hawblitzel}, {and} \bibinfo{person}{Galen Hunt}.}
  \bibinfo{year}{2009}\natexlab{}.
\newblock \showarticletitle{Helios: Heterogeneous Multiprocessing with
  Satellite Kernels}. In \bibinfo{booktitle}{\emph{ACM Symposium on Operating
  Systems Principles}}. \bibinfo{publisher}{ACM}, \bibinfo{address}{Big Sky,
  MT, US}, \bibinfo{pages}{221--234}.
\newblock


\bibitem[\protect\citeauthoryear{Nipkow, Paulson, and Wenzel}{Nipkow
  et~al\mbox{.}}{2002}]%
        {Nipkow_PW:Isabelle}
\bibfield{author}{\bibinfo{person}{Tobias Nipkow}, \bibinfo{person}{Lawrence
  Paulson}, {and} \bibinfo{person}{Markus Wenzel}.}
  \bibinfo{year}{2002}\natexlab{}.
\newblock \bibinfo{booktitle}{\emph{{Isabelle/HOL} --- A Proof Assistant for
  Higher-Order Logic}}. \bibinfo{series}{Lecture Notes in Computer Science},
  Vol.~\bibinfo{volume}{2283}.
\newblock \bibinfo{publisher}{Springer}.
\newblock
\urldef\tempurl%
\url{https://doi.org/10.1007/3-540-45949-9}
\showDOI{\tempurl}


\bibitem[\protect\citeauthoryear{Osvik, Shamir, and Tromer}{Osvik
  et~al\mbox{.}}{2006}]%
        {Osvik_ST_06}
\bibfield{author}{\bibinfo{person}{Dag~Arne Osvik}, \bibinfo{person}{Adi
  Shamir}, {and} \bibinfo{person}{Eran Tromer}.}
  \bibinfo{year}{2006}\natexlab{}.
\newblock \showarticletitle{Cache Attacks and Countermeasures: The Case of
  {AES}}. In \bibinfo{booktitle}{\emph{Proceedings of the 2006 Crytographers'
  track at the RSA Conference on Topics in Cryptology}}.
  \bibinfo{publisher}{Springer}, \bibinfo{address}{San Jose, CA, US},
  \bibinfo{pages}{1--20}.
\newblock


\bibitem[\protect\citeauthoryear{Page}{Page}{2005}]%
        {Page2005Partition}
\bibfield{author}{\bibinfo{person}{Daniel Page}.}
  \bibinfo{year}{2005}\natexlab{}.
\newblock \showarticletitle{Partitioned Cache Architecture as a Side-Channel
  Defence Mechanism}.
\newblock \bibinfo{journal}{\emph{IACR Cryptology ePrint Archive}}
  \bibinfo{volume}{2005} (\bibinfo{year}{2005}), \bibinfo{pages}{280}.
\newblock


\bibitem[\protect\citeauthoryear{Percival}{Percival}{2005}]%
        {Percival_05}
\bibfield{author}{\bibinfo{person}{Colin Percival}.}
  \bibinfo{year}{2005}\natexlab{}.
\newblock \showarticletitle{Cache Missing for Fun and Profit}. In
  \bibinfo{booktitle}{\emph{BSDCan 2005}}. \bibinfo{address}{Ottawa, CA}, 13.
\newblock
\urldef\tempurl%
\url{http://css.csail.mit.edu/6.858/2014/readings/ht-cache.pdf}
\showURL{%
\tempurl}


\bibitem[\protect\citeauthoryear{Quisquater and Samyde}{Quisquater and
  Samyde}{2001}]%
        {Quisquater_Samyde_01}
\bibfield{author}{\bibinfo{person}{Jean-Jacques Quisquater} {and}
  \bibinfo{person}{David Samyde}.} \bibinfo{year}{2001}\natexlab{}.
\newblock \showarticletitle{Electromagnetic Analysis ({EMA}): Measures and
  Counter-measures for Smart Cards}. In \bibinfo{booktitle}{\emph{Smart Card
  Programming and Security (E-Smart'01)}}. \bibinfo{address}{Cannes, FR},
  \bibinfo{pages}{200--210}.
\newblock


\bibitem[\protect\citeauthoryear{{RISC-V International}}{{RISC-V
  International}}{2023}]%
        {uSC-sig}
\bibfield{author}{\bibinfo{person}{{RISC-V International}}.}
  \bibinfo{year}{2023}\natexlab{}.
\newblock \bibinfo{title}{Microarchitecture Side Channels Special Interest
  Group {(uSC SIG)}}.
\newblock
\newblock
\urldef\tempurl%
\url{https://github.com/riscv-admin/uarch-side-channels}
\showURL{%
\tempurl}


\bibitem[\protect\citeauthoryear{Schaefer, Gold, Linde, and Scheid}{Schaefer
  et~al\mbox{.}}{1977}]%
        {Schaefer_GLS_77}
\bibfield{author}{\bibinfo{person}{Marvin Schaefer}, \bibinfo{person}{Barry
  Gold}, \bibinfo{person}{Richard Linde}, {and} \bibinfo{person}{John Scheid}.}
  \bibinfo{year}{1977}\natexlab{}.
\newblock \showarticletitle{Program Confinement in {KVM/370}}. In
  \bibinfo{booktitle}{\emph{Proceedings of the Annual ACM Conference}}.
  \bibinfo{publisher}{ACM}, \bibinfo{address}{Atlanta, GA, US},
  \bibinfo{pages}{404--410}.
\newblock
\urldef\tempurl%
\url{https://doi.org/10.1145/800179.1124633}
\showDOI{\tempurl}


\bibitem[\protect\citeauthoryear{{seL4 Foundation}}{{seL4 Foundation}}{2021}]%
        {rv64_infoflow}
\bibfield{author}{\bibinfo{person}{{seL4 Foundation}}.}
  \bibinfo{year}{2021}\natexlab{}.
\newblock \bibinfo{title}{Proof that {seL4} enforces confidentiality
  established for {RISC-V}}.
\newblock
\newblock
\urldef\tempurl%
\url{https://sel4.systems/news/2021#risc-v-infoflow}
\showURL{%
\tempurl}


\bibitem[\protect\citeauthoryear{{seL4 Foundation}}{{seL4 Foundation}}{2023}]%
        {github:l4v}
\bibfield{author}{\bibinfo{person}{{seL4 Foundation}}.}
  \bibinfo{year}{2023}\natexlab{}.
\newblock \bibinfo{title}{{seL4} Proofs}.
\newblock
\newblock
\urldef\tempurl%
\url{https://github.com/seL4/l4v}
\showURL{%
\tempurl}


\bibitem[\protect\citeauthoryear{Sewell, Myreen, and Klein}{Sewell
  et~al\mbox{.}}{2013}]%
        {Sewell_MK_13}
\bibfield{author}{\bibinfo{person}{Thomas Sewell}, \bibinfo{person}{Magnus
  Myreen}, {and} \bibinfo{person}{Gerwin Klein}.}
  \bibinfo{year}{2013}\natexlab{}.
\newblock \showarticletitle{Translation Validation for a Verified {OS} Kernel}.
  In \bibinfo{booktitle}{\emph{ACM SIGPLAN Conference on Programming Language
  Design and Implementation}}. \bibinfo{publisher}{ACM},
  \bibinfo{address}{Seattle, Washington, USA}, \bibinfo{pages}{471--481}.
\newblock


\bibitem[\protect\citeauthoryear{Sison, Buckley, Murray, Klein, and
  Heiser}{Sison et~al\mbox{.}}{2023}]%
        {Sison_BMKH_23}
\bibfield{author}{\bibinfo{person}{Robert Sison}, \bibinfo{person}{Scott
  Buckley}, \bibinfo{person}{Toby Murray}, \bibinfo{person}{Gerwin Klein},
  {and} \bibinfo{person}{Gernot Heiser}.} \bibinfo{year}{2023}\natexlab{}.
\newblock \showarticletitle{Formalising the Prevention of Microarchitectural
  Timing Channels by Operating Systems}. In
  \bibinfo{booktitle}{\emph{International Symposium on Formal Methods (FM)}}.
  \bibinfo{publisher}{Springer}, \bibinfo{address}{L\"{u}beck, DE}, 19.
\newblock
\urldef\tempurl%
\url{https://doi.org/10.1007/978-3-031-27481-7_8}
\showDOI{\tempurl}


\bibitem[\protect\citeauthoryear{Tuch, Klein, and Norrish}{Tuch
  et~al\mbox{.}}{2007}]%
        {Tuch_KN_07}
\bibfield{author}{\bibinfo{person}{Harvey Tuch}, \bibinfo{person}{Gerwin
  Klein}, {and} \bibinfo{person}{Michael Norrish}.}
  \bibinfo{year}{2007}\natexlab{}.
\newblock \showarticletitle{Types, Bytes, and Separation Logic}. In
  \bibinfo{booktitle}{\emph{ACM SIGPLAN-SIGACT Symposium on Principles of
  Programming Languages}}. \bibinfo{publisher}{ACM}, \bibinfo{address}{Nice,
  France}, \bibinfo{pages}{97--108}.
\newblock


\bibitem[\protect\citeauthoryear{Wang, Ferraiuolo, and Suh}{Wang
  et~al\mbox{.}}{2014}]%
        {Wang_FS_14}
\bibfield{author}{\bibinfo{person}{Yao Wang}, \bibinfo{person}{Andrew
  Ferraiuolo}, {and} \bibinfo{person}{G.~Edward Suh}.}
  \bibinfo{year}{2014}\natexlab{}.
\newblock \showarticletitle{Timing channel protection for a shared memory
  controller}. In \bibinfo{booktitle}{\emph{Proceedings of the 20th IEEE
  Symposium on High-Performance Computer Architecture}}.
  \bibinfo{address}{Orlando, FL, US}.
\newblock


\bibitem[\protect\citeauthoryear{Wang and Lee}{Wang and Lee}{2007}]%
        {Wang_Lee_07}
\bibfield{author}{\bibinfo{person}{Zhenghong Wang} {and}
  \bibinfo{person}{Ruby~B. Lee}.} \bibinfo{year}{2007}\natexlab{}.
\newblock \showarticletitle{New Cache Designs for Thwarting Software
  Cache-based Side Channel Attacks}. In \bibinfo{booktitle}{\emph{Proceedings
  of the 34th International Symposium on Computer Architecture}}.
  \bibinfo{publisher}{ACM}, \bibinfo{address}{San Diego, CA, US},
  \bibinfo{pages}{494--505}.
\newblock


\bibitem[\protect\citeauthoryear{Wang and Lee}{Wang and Lee}{2008}]%
        {Wang_Lee_08}
\bibfield{author}{\bibinfo{person}{Zhenghong Wang} {and}
  \bibinfo{person}{Ruby~B. Lee}.} \bibinfo{year}{2008}\natexlab{}.
\newblock \showarticletitle{A novel cache architecture with enhanced
  performance and security}. In \bibinfo{booktitle}{\emph{Proceedings of the
  41st ACM/IEE International Symposium on Microarchitecture}}.
  \bibinfo{address}{Lake Como, Italy}, \bibinfo{pages}{83--93}.
\newblock


\bibitem[\protect\citeauthoryear{Wistoff, Schneider, G\"{u}rkaynak, Benini, and
  Heiser}{Wistoff et~al\mbox{.}}{2021}]%
        {Wistoff_SGBH_21}
\bibfield{author}{\bibinfo{person}{Nils Wistoff}, \bibinfo{person}{Moritz
  Schneider}, \bibinfo{person}{Frank G\"{u}rkaynak}, \bibinfo{person}{Luca
  Benini}, {and} \bibinfo{person}{Gernot Heiser}.}
  \bibinfo{year}{2021}\natexlab{}.
\newblock \showarticletitle{Microarchitectural Timing Channels and their
  Prevention on an Open-Source 64-bit {RISC}-{V} Core}. In
  \bibinfo{booktitle}{\emph{Design, Automation and Test in Europe (DATE)}}.
  \bibinfo{publisher}{IEEE}, \bibinfo{address}{virtual}, 6.
\newblock


\bibitem[\protect\citeauthoryear{Wistoff, Schneider, G\"{u}rkaynak, Heiser, and
  Benini}{Wistoff et~al\mbox{.}}{2023}]%
        {Wistoff_SGHB_23}
\bibfield{author}{\bibinfo{person}{Nils Wistoff}, \bibinfo{person}{Moritz
  Schneider}, \bibinfo{person}{Frank G\"{u}rkaynak}, \bibinfo{person}{Gernot
  Heiser}, {and} \bibinfo{person}{Luca Benini}.}
  \bibinfo{year}{2023}\natexlab{}.
\newblock \showarticletitle{Systematic Prevention of On-Core Timing Channels by
  Full Temporal Partitioning}.
\newblock \bibinfo{journal}{\emph{IEEE Trans. Comput.}} \bibinfo{volume}{72},
  \bibinfo{number}{5} (\bibinfo{year}{2023}), \bibinfo{pages}{1420--1430}.
\newblock
\urldef\tempurl%
\url{https://doi.org/10.1109/TC.2022.3212636}
\showDOI{\tempurl}


\bibitem[\protect\citeauthoryear{Wray}{Wray}{1991}]%
        {Wray_91}
\bibfield{author}{\bibinfo{person}{John~C. Wray}.}
  \bibinfo{year}{1991}\natexlab{}.
\newblock \showarticletitle{An analysis of covert timing channels}. In
  \bibinfo{booktitle}{\emph{Proceedings of the 1991 IEEE Computer Society
  Symposium on Research in Security and Privacy}}. \bibinfo{publisher}{IEEE},
  \bibinfo{address}{Oakland, CA, US}, \bibinfo{pages}{2--7}.
\newblock


\bibitem[\protect\citeauthoryear{Wu, Xu, and Wang}{Wu et~al\mbox{.}}{2012}]%
        {Wu_XW_12}
\bibfield{author}{\bibinfo{person}{Zhenyu Wu}, \bibinfo{person}{Zhang Xu},
  {and} \bibinfo{person}{Haining Wang}.} \bibinfo{year}{2012}\natexlab{}.
\newblock \showarticletitle{Whispers in the Hyper-space: High-speed Covert
  Channel Attacks in the Cloud}. In \bibinfo{booktitle}{\emph{Proceedings of
  the 21st USENIX Security Symposium}}. \bibinfo{publisher}{USENIX},
  \bibinfo{address}{Bellevue, WA, US}, \bibinfo{pages}{159--173}.
\newblock


\bibitem[\protect\citeauthoryear{Zellweger, Gerber, Kourtis, and
  Roscoe}{Zellweger et~al\mbox{.}}{2014}]%
        {Zellweger_GKR_14}
\bibfield{author}{\bibinfo{person}{Gerd Zellweger}, \bibinfo{person}{Simon
  Gerber}, \bibinfo{person}{Kornilios Kourtis}, {and} \bibinfo{person}{Timothy
  Roscoe}.} \bibinfo{year}{2014}\natexlab{}.
\newblock \showarticletitle{Decoupling Cores, Kernels, and Operating Systems}.
  In \bibinfo{booktitle}{\emph{USENIX Symposium on Operating Systems Design and
  Implementation}}. \bibinfo{publisher}{USENIX}, \bibinfo{address}{Broomfield,
  CO, US}, \bibinfo{pages}{17--31}.
\newblock


\end{thebibliography}
  \fi
}
\end{document}